\documentclass[useAMS,usenatbib]{mn2e}

\usepackage[english]{babel}
\usepackage[ansinew]{inputenc}
\usepackage{amssymb,amsmath}
\usepackage{graphicx} 
\usepackage{natbib} 
\usepackage{wrapfig}
\usepackage{color}
\usepackage{ulem}
\usepackage{ae}

\newcommand{\prob}{\textrm{prob}}

\newcommand{\Mpc}{\textrm{Mpc}}
\newcommand{\kms}{\textrm{km}/\textrm{s}}

\title[Only marginal alignment of disc galaxies]{Only marginal alignment of disc galaxies}

\author[R. Andrae \& K. Jahnke (2011)]{Ren\'e Andrae$^{1}$\thanks{E-mail:
andrae@mpia-hd.mpg.de} and Knud Jahnke$^{1}$\\
$^{1}$Max-Planck-Institut f\"ur Astronomie, K\"onigstuhl 17, 69117 Heidelberg, Germany}

\begin{document}

\date{Received 2011 June 10. Accepted 2011 August 11.}

\pagerange{\pageref{firstpage}--\pageref{lastpage}} \pubyear{2011}

\maketitle

\label{firstpage}

\begin{abstract}
Testing theories of angular-momentum acquisition of rotationally supported disc galaxies is the key to understand the formation of this type of galaxies. The tidal-torque theory tries to explain this acquisition process in a cosmological framework and predicts positive autocorrelations of angular-momentum orientation and spiral-arm handedness, i.e., alignment of disc galaxies, on short distance scales of 1Mpc/h. This disc alignment can also cause systematic effects in weak-lensing measurements. Previous observations claimed discovering these correlations but are overly optimistic in the reported level of statistical significance of the detections. Errors in redshift, ellipticity and morphological classifications were not taken into account, although they have a significant impact. We explain how to rigorously propagate all important errors through the estimation process. Analysing disc galaxies in the SDSS database, we find that positive autocorrelations of spiral-arm handedness and angular-momentum orientations on distance scales of 1Mpc/h are plausible but not statistically significant. Current data appears not good enough to constrain parameters of theory. This result agrees with a simple hypothesis test in the Local Group, where we also find no evidence for disc alignment. Moreover, we demonstrate that ellipticity estimates based on second moments are strongly biased by galactic bulges even for Scd galaxies, thereby corrupting correlation estimates and overestimating the impact of disc alignment on weak-lensing studies. Finally, we discuss the potential of future sky surveys. We argue that photometric redshifts have too large errors, i.e., PanSTARRS and LSST cannot be used. Conversely, the EUCLID project will not cover the relevant redshift regime. We also discuss potentials and problems of front-edge classifications of galaxy discs in order to improve autocorrelation estimates of angular-momentum orientation.
\end{abstract}



\section{Introduction}

Disc galaxies constitute a substantial part of the galaxy population in the nearby universe \citep{Bamford2009}. As these galaxies are rotationally supported, it is of vital importance to understand how disc galaxies acquire their angular momentum. The tidal-torque theory tries to explain this angular-momentum acquisition through tidal shearing from the dark-matter host halo's gravitational field and the moment of inertia of the forming protogalaxy \citep[for a recent review see][]{Schaefer2009}. This theory predicts alignment effects of disc galaxies, since angular-momentum acquisition is partially governed by environmental effects such that neighbouring disc galaxies residing in the same environment should exhibit similar angular momenta. Hence, testing intrinsic alignments of angular momenta of disc galaxies provides a fundamental test for our understanding of galaxy formation in the cosmological framework. Apart from enhancing our understanding of disc-galaxy formation, investigating these alignment effects is also important because they constitute a potentially significant systematic effect in weak-gravitational-lensing surveys \citep[e.g.][]{Crittenden2001}.

For this goal, we use autocorrelation estimates of spiral-arm handedness and galactic angular-momentum-orientation vectors, respectively. We revisit the works by \citet{Slosar2009} and \citet{Lee2010} and explain that these estimations do not take into account all relevant error contributions and are therefore too optimistic in the reported statistical significance. In this article, we explain how to incorporate the relevant error sources and demonstrate their impact on the results. This methodological rigour is also in a general sense highly relevant, since at the frontier of astrophysical research data analysis can otherwise produce misleading results. Typically for methodological studies, the basic principle and the techniques presented here are also applicable to other astrophysical investigations which involve the estimation of spatial two-point correlation functions, for instances, investigations of baryonic acoustic oscillations (BAOs) \citep[e.g.][]{Blake2011}. Although estimators used for investigations of BAOs are usually much more elaborate than the simple estimator we are going to use, our assessment of impact of (star-galaxy) classification and redshift errors also applies to this setting.\footnote{The perfect BAO estimator would be a generative model that for every galaxy predicts the redshift and star-galaxy classification probability based on the observation conditions. This would enable us to directly take into account these error sources.} However, we also go beyond the purely methodological aspect and discuss the potential of improving the autocorrelation estimates with new surveys in order to eventually obtain astrophysical results. In particular, we discuss the potential of estimating the front edges of disc galaxies via dust extinction in order to improve correlation estimates of angular-momentum orientation.

\subsection{Strategy}

We start in Sect.~\ref{sect:local_group} by investigating the orientations of angular-momentum vectors in the Local Group. This is meant as an exercise, motivating the necessity of correlation functions. We then present in Sect.~\ref{sect:data} the details and selection criteria of the data samples we are using. In Sect.~\ref{sect:Correlation_estimators}, we explain how to obtain correlation estimates and their corresponding error estimates for both handedness and angular-momentum-orientation vectors. The main body of this article is Sect.~\ref{sect:impact_errors}, where we explain the difference between conditional and marginal errors, discuss the relevant error contributions, and explain how to propgate errors numerically by simple Monte-Carlo sampling. In that section, we estimate marginal autocorrelations of handedness and angular-momentum orientations, respectively. This is also the section relevant to readers who are interested in marginal estimates of correlation functions in general, e.g., in the context of baryonic acoustic oscillations. In an attempt to improve the statistical significance of our results by replacing isophotal ellipticity estimates by less noisy estimators, we show in Sect.~\ref{sect:eps_biases_2nd_moments} that ellipticities based on second moments are strongly biased. We clearly demonstrate that this bias corrupts correlation estimates. We outline possible improvements and the potential of future sky surveys in Sect.~\ref{sect:future_surveys}. We discuss our final results and conclude in Sect.~\ref{sect:conclusions}.

\section{Are angular momenta randomly oriented in the Local Group?}
\label{sect:local_group}

As we are investigating the alignment of angular momenta of disc galaxies, the Local Group is a natural first testbed. Apart from numerous dwarf galaxies, the Local Group consists of four disc galaxies, namely the Milky Way, Andromeda (M31), M33, and the Large Magellanic Cloud (LMC), all with pairwise distances of less than 1Mpc.

\subsection{Angular-momentum orientation of the Milky Way}

We start by estimating the angular-momentum-orientation vector of the Milky Way in equatorial coordinates. In order to estimate the angular-momentum-orientation vector of the Milky Way, we need two ingredients:
\begin{enumerate}
\item The unit vector $\vec r_\odot$ pointing from the Galactic centre to the position of the Sun.
\item The unit vector $\vec v_\odot$ of the Sun's velocity on its trajectory around the Galactic centre.
\end{enumerate}
Given the valid assumption that the Sun lies inside and is co-rotating with the Galactic disc, we can then compute the Milky Way's angular-momentum-orientation vector via
\begin{equation}\label{eq:L_from_r_and_v}
\vec L_\textrm{MW}=\vec r_\odot\times\vec v_\odot \;\textrm{.}
\end{equation}
We can infer $\vec r_\odot$ from the equatorial coordinates of the Galactic centre, $\alpha_\textrm{GC}\approx 266.42^\circ$ and $\delta_\textrm{GC}\approx -29.01^\circ$. Here, we have to keep in mind that these coordinates are pointing from the Sun towards the Galactic centre, i.e., $\vec r_\odot$ is the inverted direction,
\begin{equation}
\vec r_\odot = -\left(\begin{array}{c}
\cos\alpha_\textrm{GC}\,\sin(90^\circ - \delta_\textrm{GC}) \\
\sin\alpha_\textrm{GC}\,\sin(90^\circ - \delta_\textrm{GC}) \\
\cos(90^\circ - \delta_\textrm{GC})
\end{array}\right) \;\textrm{.}
\end{equation}
The unit vector $\vec v_\odot$ has to be inferred from the rotation of the Galactic disc. By definition, $\vec v_\odot$ points into the direction specified by Galactic longitude $\ell=90^\circ$ and Galactic latitude $b=0^\circ$ \citep[e.g.][]{Brunthaler2005}. In equatorial coordinates, this direction is given by $\alpha_v\approx 318.00^\circ$ and $\delta_v\approx 48.33^\circ$, such that
\begin{equation}
\vec v_\odot = \left(\begin{array}{c}
\cos\alpha_v\,\sin(90^\circ - \delta_v) \\
\sin\alpha_v\,\sin(90^\circ - \delta_v) \\
\cos(90^\circ - \delta_v)
\end{array}\right) \;\textrm{.}
\end{equation}
Inserting these values into Eq.~(\ref{eq:L_from_r_and_v}), we obtain the following estimate of the angular-momentum-orientation vector of the Milky Way,
\begin{equation}\label{eq:angular_momentum_vector_MilkyWay}
\vec L_\textrm{MW} \approx \left(\begin{array}{c}
\phantom{-}0.86771 \\
\phantom{-}0.19878 \\
-0.45560
\end{array}\right) \;\textrm{.}
\end{equation}
We conduct two cross-checks: First, the two unit vectors $\vec r_\odot$ and $\vec v_\odot$ should be orthogonal and indeed their scalar product is $\vec r_\odot\cdot\vec v_\odot\approx 0.00097\ll 1$. Second, $\vec L_\textrm{MW}$ by construction should be normal to the plane of the Galactic disc, i.e., parallel or antiparallel to the unit vector pointing into the direction of the Galactic North pole, whose equatorial coordinates are given by $\alpha_\textrm{NP}\approx 192.86^\circ$ and $\delta_\textrm{NP}\approx 27.13^\circ$. Indeed, the scalar product is $\vec L_\textrm{MW}\cdot\vec u_\textrm{NP}\approx -0.9999992$, i.e., both vectors are almost perfectly antiparallel.

\subsection{Angular-momentum orientations of Andromeda, M33, and the LMC}
\label{sect:L_of_M31_M33}

In order to estimate the angular-momentum orientations of Andromeda, M33, and the LMC, we use the formalism described in \citet{Lee2010} which is based on ellipticity estimates and the assumption of intrinsically circular galactic discs.

\subsubsection{Andromeda (M31)}

For Andromeda, we adopt an inclination angle of $77^\circ$ \citep{Walterbos1988} and an orientation angle of $38^\circ$ \citep{Walterbos1987}. Furthermore, dust lanes enable us to identify the front edge of Andromeda's galactic disc, which is the North-Western edge. Our front-edge estimate agrees with the result of \citet{Iye1999} who investigated the reddening of globular clusters as a function of height above the major axis. Given its equatorial coordinates $\alpha_\textrm{M31}\approx 10.69^\circ$ and $\delta_\textrm{M31}\approx 41.27^\circ$, we can compute the angular-momentum-orientation vector of Andromeda up to its sign. \citet{Chemin2009} published spatially resolved HI spectra of M31, which enable us to infer the disc rotation directly. Their map of radial velocities directly implies that the North-Eastern part is receding from us, whereas the South-Western part is rotating towards us. Consequently, the angular-momentum-orientation vector of Andromeda points South-East and away from our own position. Therefore, if we project $\vec L_\textrm{M31}$ onto the unit direction vector pointing from the Milky Way towards Andromeda, this projection must be positive. This condition enables us to fully determine the angular-momentum-orientation vector of Andromeda,
\begin{equation}\label{eq:angular_momentum_vector_M31}
\vec L_\textrm{M31} \approx \left(\begin{array}{c}
          -0.08031 \\
          -0.79651 \\
\phantom{-}0.59926
\end{array}\right) \;\textrm{.}
\end{equation}

\subsubsection{Triangulum Galaxy (M33)}

Concerning M33, we adopt an inclination angle of $49^\circ$ and an orientation angle of $21^\circ$ \citep{Corbelli1997}. M33 clearly is a right-handed (Z-wise) spiral. This rotational sense agrees with the results of \citet{Brunthaler2005} who observed the proper motion of two H${}_2$O masers in M33. It also agrees with the results of \citet{Putman2009}, who measured the radial-velocity field of HI gas in M33.  Again, this implies that the projections of both possible front-edge configurations of $\vec L_\textrm{M33}$ onto the unit direction vector pointing from the Milky Way towards M33 have to be positive. Unfortunately, M33 does not exhibit dust lanes, such that the front edge remains unknown. This is not surprising since M33 is not as highly inclined as Andromeda such that we are less likely to observe a dust lane. From dust reddening of C-rich AGB stars \citet{Cioni2008} concluded that there is weak evidence that the North-Western side of M33 is the front-edge. Given its equatorial coordinates $\alpha_\textrm{M33}\approx 23.46^\circ$ and $\delta_\textrm{M33}\approx 30.66^\circ$, the angular-momentum-orientation vector of M33 then reads
\begin{equation}\label{eq:angular_momentum_vector_M33}
\vec L_\textrm{M33} \approx \left(\begin{array}{c}
\phantom{-}0.67170 \\
          -0.47655 \\
\phantom{-}0.56721
\end{array}\right) \,\textrm{.}
\end{equation}
The front-edge estimate of \citet{Cioni2008} is still rather uncertain (see their Fig.~9). However, it is sufficient for this exercise.

\subsubsection{Large Magellanic Cloud (LMC)}

Concerning the LMC, we adopt an inclination angle of $35^\circ$ and an orientation angle of $123^\circ$ \citep{Marel2001}.  Furthermore, \citet{Marel2001} find clear evidence that the North-Eastern side of the disc is the front-edge (their Fig.~5). The rotational sense of the LMC is right-handed as is evident from observed velocity fields \citep[e.g.][]{Olsen2007}. Given its equatorial coordinates $\alpha_\textrm{LMC}\approx 80.8938^\circ$ and $\delta_\textrm{LMC}\approx -69.7561^\circ$, the angular-momentum-orientation vector of the LMC then reads
\begin{equation}\label{eq:angular_momentum_vector_LMC}
\vec L_\textrm{LMC} \approx \left(\begin{array}{c}
-0.29699 \\
-0.46945 \\
-0.83152
\end{array}\right) \,\textrm{.}
\end{equation}

\subsection{Random orientation}
\label{sect:local_group_hypothesis_test}

\begin{figure}
\includegraphics[width=8cm]{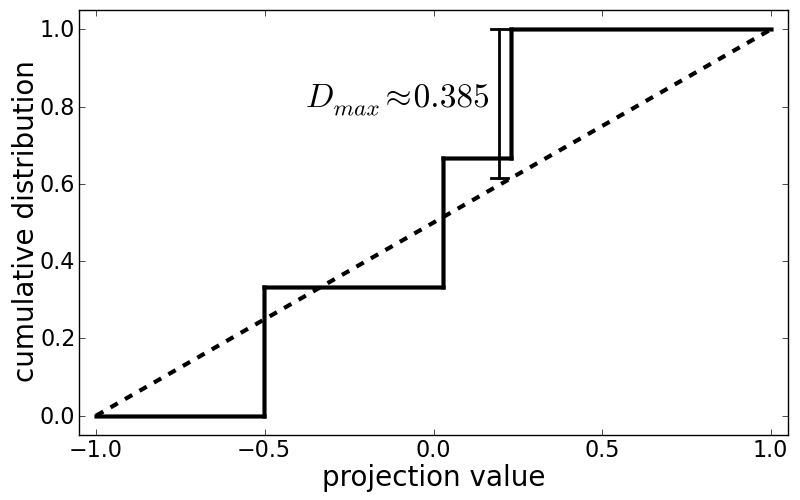}
\caption{KS-test of angular-momentum-orientation vectors in the Local Group. Step function: Empirical (unbiased) estimate of cumulative distribution for the Local Group. Dashed line: Cumulative distribution of null hypothesis of random orientations. The maximum KS-distance is $D_\textrm{max}\approx 0.385$.}
\label{fig:KSTest_local_group}
\end{figure}

Are the angular-momentum-orientation vectors in the Local Group compatible with the null hypothesis of random orientation? In order to test this, we investigate the distribution of projection values. For the four disc galaxies, there can only derive three statistically independent projections. We choose the projections onto the Milky Way:
\begin{itemize}
\item $\vec L_\textrm{MW}\cdot\vec L_\textrm{M31}\approx -0.5010$
\item $\vec L_\textrm{MW}\cdot\vec L_\textrm{M33}\approx +0.2297$
\item $\vec L_\textrm{MW}\cdot\vec L_\textrm{LMC}\approx +0.0278$
\end{itemize}
Adding further projection values, e.g., $\vec L_\textrm{M31}\cdot\vec L_\textrm{M33}$, would introduce correlations compromising the KS-test. Figure~\ref{fig:KSTest_local_group} shows the resulting cumulative distribution of projection values for the Local Group. Furhtermore, Fig.~\ref{fig:KSTest_local_group} shows the cumulative distribution for the null hypothesis where all projection values are equally likely. The KS-distance is then $D_\textrm{max}\approx 0.385$ which yields a $p$-value of $\approx 0.648$ \citep{Press2002}. Consequently, with 64.8\% probability we make a mistake if we reject the null hypothesis of randomly oriented angular-momentum-orientation vectors in the Local Group.

We conclude from this simple hypothesis test that there is no evidence that disc alignment is at work in the Local Group. However, this hypothesis test is rather coarse given the small number of disc galaxies and the neglection of galaxy separations. Hence, a more elaborate investigation naturally leads us to spatial autocorrelation functions estimated from large samples of disc galaxies as the key diagnostic tool for investigations of disc alignment.

\section{The Data}
\label{sect:data}

An autocorrelation analysis of angular momenta requires a survey covering a large area with homogeneous galaxy morphologies in order to (a) select disc galaxies and (b) estimate their three-dimensional angular-momentum-orientation vectors. The best database for this purpose is the SDSS. We exploit visual morphological classifications from the Galaxy Zoo project and automated classifications from \citet{HuertasCompany2011}, enhanced by additional information from the general SDSS database.

\subsection{Galaxy Zoo}
\label{sect:about_galaxy_zoo}

Galaxy Zoo \citep{Lintott2008,Bamford2009,Lintott2011} is a unique project where the morphology of nearly 900,000 galaxies from the Sloan Digital Sky Survey (SDSS) spectroscopic sample have been classified visually by the internet community. Each galaxy has been classified multiple times by different internet users, which provides a probabilistic object-to-class assignment. Concerning galaxy morphologies, such a probabilistic assignment is more physical than a hard assignment, as has been discussed by \citet{Andrae2010a}. In detail, the Galaxy Zoo database provides probabilistic assignments to the following morphological classes:
\begin{itemize}
\item elliptical, $p_\textrm{ell}^\textrm{GZ}$,
\item disc, $p_\textrm{disc}^\textrm{GZ}$,
\item edge-on disc, $p_\textrm{edge}^\textrm{GZ}$,
\item clock-wise/Z-wise spiral in projection, $p_\textrm{Z}^\textrm{GZ}$,
\item anti-clock-wise/S-wise spiral in projection, $p_\textrm{S}^\textrm{GZ}$,
\item merger, $p_\textrm{mg}^\textrm{GZ}$.
\end{itemize}
All probabilities that are taken from Galaxy Zoo carry a ``GZ'' superscript. The normalisation is given by
\begin{equation}\label{eq:GalZoo_prob_normalisation}
p_\textrm{ell}^\textrm{GZ} + p_\textrm{disc}^\textrm{GZ} + p_\textrm{edge}^\textrm{GZ} + p_\textrm{Z}^\textrm{GZ} + p_\textrm{S}^\textrm{GZ} + p_\textrm{mg}^\textrm{GZ} = 1 \,\textrm{.}
\end{equation}

\citet{Land2008} reported a bias in the handedness classifications, $p_\textrm{Z}^\textrm{GZ}$ and $p_\textrm{S}^\textrm{GZ}$ , where more spiral galaxies are classified as S-wise than as Z-wise.\footnote{\citet{Land2008} also used flipped galaxy images and still observed an excess of S-wise over Z-wise spirals in visual classifications. The exact origin of this bias is unknown, though one option considered by \citet{Land2008} is a psychological effect.} This bias is corrected in an asymmetric, additive fashion by \citet{Land2008} and \citet{Slosar2009} in order to enforce that the proportions of Z-wise and S-wise spirals are equal with regard to the whole sample. In contrast to this, we employ a \textit{symmetric}, additive bias correction of the form
\begin{equation}\label{eq:handedness_bias_correction}
p_\textrm{Z} = p_\textrm{Z}^\textrm{GZ} + b \qquad\textrm{and}\qquad p_\textrm{S} = p_\textrm{S}^\textrm{GZ} - b \,\textrm{,}
\end{equation}
where $b$ is chosen such that the numbers of Z-wise and S-wise spirals are identical. There are two reasons:
\begin{enumerate}
\item The symmetric correction preserves the normalisation of Eq.~(\ref{eq:GalZoo_prob_normalisation}). This is important because in contrast to \citet{Slosar2009} we are handling the Galaxy Zoo results fully probabilistically in our analysis (cf.\ Sect.~\ref{sect:uncertainties_classification}).
\item Demanding that the proportions of Z-wise and S-wise spirals are equal only provides a single condition, such that an asymmetric correction with two biases, $b_\textrm{Z}$ and $b_\textrm{S}$, is not fully constrained and therefore arbitrary.
\end{enumerate}
Our value of $b$ is 0.0105 and thus similar to \citet{Land2008}. \citet{Slosar2009} argued that such a bias can only lead to a constant offset in the handedness autocorrelation function, but it cannot feign a distance-dependent autocorrelation, which is the predicted astrophysical signal.

\subsection{Catalogue of \citet{HuertasCompany2011}}

Similar to the Galaxy Zoo project, \citet{HuertasCompany2011} performed a morphological classification on the SDSS spectroscopic galaxy sample. There are two important differences with respect to Galaxy Zoo:
\begin{enumerate}
\item The morphological classes are:
\begin{itemize}
\item elliptical, $p_\textrm{ell}^\textrm{HC}$,
\item S0 galaxy, $p_\textrm{S0}^\textrm{HC}$,
\item Sab disc galaxy, $p_\textrm{Sab}^\textrm{HC}$,
\item Scd disc galaxy, $p_\textrm{Scd}^\textrm{HC}$.
\end{itemize}
All probabilities taken from the catalogue of \citet{HuertasCompany2011} carry a superscript ``HC''.
\item Instead of visual inspection, a support-vector machine, i.e., an automated classification algorithm, has been used in order to classify the galaxies.
\end{enumerate}
The normalisation reads
\begin{equation}
p_\textrm{ell}^\textrm{HC} + p_\textrm{S0}^\textrm{HC} + p_\textrm{Sab}^\textrm{HC} + p_\textrm{Scd}^\textrm{HC} = 1 \;\textrm{.}
\end{equation}
As mentioned in \citet{HuertasCompany2011}, Sect.~3.1 therein, the ``Scd'' class not only contains Scd galaxies but also irregular galaxies.

\subsection{Additional information from the SDSS database}

The Galaxy Zoo catalogue and the catalogue of \citet{HuertasCompany2011} have been cross-matched with the general SDSS database via the spectral object IDs (\textsc{SpecObjID}) of the galaxies. We exploit this matching in order to obtain additional information about the visually classified galaxies. In particular, we retrieved the following information from the SDSS database:
\begin{itemize}
\item $r$-band ellipticity:
\begin{itemize}
\item isophotal axis ratio and orientation angle,
\item Stokes parameters and including their errors,
\end{itemize}
\item (circular) Petrosian radii in $r$- and $i$-band,
\item spectroscopic redshift estimate and its error.
\end{itemize}
The two Stokes parameters $Q$ and $U$ encode the complex ellipticity \citep[e.g.][]{Bartelmann2001}
\begin{equation}
\epsilon = \epsilon_+ + i\epsilon_\times = Q + iU = \frac{1-q}{1+q}e^{2i\theta} \;\textrm{,}
\end{equation}
where $q = b/a$ denotes the ratio of semi-minor over semi-major axis and $\theta$ denotes the orientation angle. From the spectroscopic redshift estimate, $\hat z$, we estimate the comoving distance,
\begin{equation}
d(\hat z) = \frac{c}{H_0}\int_0^{\hat z} \frac{dz}{(1+z)^2\sqrt{(1+z)^3\Omega_m + \Omega_\Lambda}} \;\textrm{,}
\end{equation}
assuming a $\Lambda$CDM cosmology with parameters $H_0 = 100h\,km\,s^{-1}Mpc^{-1}$, $\Omega_\Lambda = 0.734$ and $\Omega_m = 1- \Omega_\Lambda$ \citep{Larson2011}. These distance estimates may suffer from peculiar motions of the galaxies \citep[see discussion in Sect.~\ref{sect:improving_redshifts} or ``Fingers-of-god'' effect in, e.g.,][]{Hamilton1998}. Using equatorial coordinates right ascension and declination angles $(\alpha,\delta)$, we convert to a three-dimensional, global, spherical coordinate system with polar angles $\varphi=\alpha$ and $\vartheta=\pi/2-\delta$. The position $\vec r$ of each galaxy is then simply given by
\begin{equation}
\vec r = d\,\vec e_r = \left(\begin{array}{c}
d\cos\varphi\sin\vartheta \\
d\sin\varphi\sin\vartheta \\
d\cos\vartheta
\end{array}\right) \;\textrm{.}
\end{equation}
Distances between two galaxies are then computed via Euclidean distances $|\vec r_1-\vec r_2|$, i.e., we assume that the Euclidean metric does not change with redshift. This is a reliable approximation, since the galaxies in our sample span a redshift range where nonlinear cosmological effects are negligible.

\subsection{Data selection}

Not all objects in the catalogue can be used for our analysis. Some objects have to be removed for various reasons. We now describe the data selection for the two galaxy samples used to estimate correlations in handedness and angular-momentum orientations.

\subsubsection{Handedness sample}

First, starting from the Galaxy Zoo sample, we select all galaxies with either $p_\textrm{Z}^\textrm{GZ}\geq 0.778$ or $p_\textrm{S}^\textrm{GZ}\geq 0.8$, which results in 36,999 galaxies. These asymmetric probability thresholds are chosen this way in order to allow for some flexibility in the correction of the handedness bias of $b=0.0105$.

Second, we obtained the $r$-band Petrosian radii from the SDSS \textsc{Galaxy} table, the spectroscopic redshift estimate and its error estimate from the SDSS \textsc{SpecObjAll} table. Actually, all objects in the Galaxy Zoo sample have been selected from the SDSS spectroscopic sample. For reasons unknown to us, we could not find 103 objects in the \textsc{Galaxy} table and another 5,106 objects were untraceable in the \textsc{SpecObjAll} table.\footnote{The Galaxy Zoo database provides the SDSS \textsc{ObjID}, which was used to identify objects in the \textsc{Galaxy} table. Cross-matching with the \textsc{SpecObjAll} table was done by retrieving the \textsc{SpecObjID} from the \textsc{Galaxy} table or -- if this label was unavailable -- by matching the given \textsc{ObjID} with the \textsc{BestObjID} from the \textsc{SpecObjAll} table.} This leaves us with 31,790 objects with $r$-band Petrosian radius and estimates of spectroscopic redshift and its error.

Third, we remove multiple objects from the sample, i.e., extended galaxies that have been shredded by the SDSS pipeline producing multiple entries of a single object. We automatically removed galaxy pairs whose angular separations were less than 1.5 times the maximum $r$-band Petrosian radius of both galaxies. Furthermore, \citet{Slosar2009} removed another 69 objects through visual inspection. This list has been kindly provided by An\v{z}e Slosar such that we are capable of removing these objects, too. This leaves us with 31,621 galaxies.

Finally, we apply the additive and symmetric bias correction of the handedness classifications given by Eq.~(\ref{eq:handedness_bias_correction}). Na\"ively interpreting any galaxy with $p_\textrm{Z}\geq 0.8-b$ as Z-wise spiral and any galaxy with $p_\textrm{S}\geq 0.8+b$ as S-wise spiral, we end up with 15,083 Z-wise and 15,071 S-wise spirals for a bias correction of $b=0.0105$. Therefore, our sample is slightly smaller than the one used by \citet{Slosar2009}.

\subsubsection{Angular-momentum-orientation sample}

Based on the catalogue of morphological classifications by \citet{HuertasCompany2011}, we select those galaxies with spectroscopic redshifts $0 < z \leq 0.02$ and probability $p_\textrm{Scd}^\textrm{HC} > 0.5$ to be a galaxy of type Sc or Sd. This leaves us with 4,236 galaxies satisfying these criteria, the same number of objects as reported by \citet{Lee2010}. For 25 of these objects we could not find any information in the SDSS database, i.e., estimates of $r$-band Petrosian radii, Stokes parameters, their errors, and error estimates of spectroscopic redshift are missing. For these objects, we set the spectroscopic redshift error to $10^{-4}$, which is a typical value for this sample. Petrosian radii are set to zero. Using the automated method described above, we find 20 rogue pairs in this sample. For each pair, we randomly discard one of the two galaxies, such that we are left with a sample of 4,216 Scd galaxies.

\subsection{From axis ratio to angular-momentum orientation}
\label{sect:from_q_to_L}

The orientation of the angular-momentum-orientation vector has to be inferred from the observed galactic disc by invoking several assumptions. We follow the formalism described, e.g., in \citet{Lee2010} in order to estimate the angular-momentum-orientation vector from the observed axis ratios, elliptical orientation angles, and equatorial coordinates. In fact, we already used this formalism in Sect.~\ref{sect:L_of_M31_M33}. If not specified otherwise, we adopt the same correction for disc thickness like \citet{Lee2010}, who assumed an intrinsic axial ratio of $p=0.1$ for Scd galaxies based on \citet{Haynes1984}. For later purposes, we note that \citet{Heidmann1972} compared different estimates of the intrinsic axial ratios and find values between $p=0.083$ and 0.145 for Scd galaxies.

\section{Correlation estimators}
\label{sect:Correlation_estimators}

In this section, we discuss the correlation estimators for angular-momentum orientations and handedness. We also explain how to estimate errors. We start by explaining the general formalism and then specialise on both angular-momentum orientations and handedness.

\subsection{Simple correlation estimator}

Given two random variates $X$ and $Y$, we want to estimate their correlation $\xi_{XY}$ and its error. If $N$ samples $x_1, x_2,\ldots, x_N$ and $y_1, y_2,\ldots, y_N$ have been drawn from $X$ and $Y$ and are independent and identically distributed, a simple correlation estimator\footnote{More elaborate estimators can be defined. Equation~(\ref{eq:xi_general}) is the maximum-likelihood estimate, if and only if $(X,Y)$ are drawn from a bivariate Gaussian, i.e., if there are no higher-order correlations.} is given by,
\begin{equation}\label{eq:xi_general}
\hat\xi_{XY} = \langle(X-\langle X\rangle)(Y-\langle Y\rangle)\rangle = \langle XY\rangle - \langle X\rangle\langle Y\rangle \;\textrm{,}
\end{equation}
where the hat on $\xi_{XY}$ indicates an estimator and
\begin{equation}
\langle XY\rangle = \frac{1}{N}\sum_{n=1}^N x_n y_n \;\textrm{,}
\end{equation}
\begin{equation}
\langle X\rangle = \frac{1}{N}\sum_{n=1}^N x_n \quad\textrm{and}\quad \langle Y\rangle = \frac{1}{N}\sum_{n=1}^N y_n \;\textrm{.}
\end{equation}
Merely obtaining a value of $\hat\xi_{XY}$ via Eq.~(\ref{eq:xi_general}) alone is not informative in any way. We also need an error estimate for $\hat\xi_{XY}$ in order to get a meaningful result. As $\hat\xi_{XY}$ is the \textit{mean} of $(X-\langle X\rangle)(Y-\langle Y\rangle)$, the variance of $\hat\xi_{XY}$ is given by the variance of $(X-\langle X\rangle)(Y-\langle Y\rangle)$ divided by $N$.\footnote{Mark the following important difference: If we are interested in estimating some random variate $Z$, we employ its mean $\langle Z\rangle$ and its variance $\langle Z^2\rangle - \langle Z\rangle^2$. However, in this case we are not interested in estimating $Z$ but in estimating the \textit{mean} of $Z$ and the variance of $\langle Z\rangle$ equals the variance of $Z$ divided by the number of samples drawn from $Z$. Loosely speaking, if we draw more samples from $Z$, the distribution of $Z$ does not change, in particular its width (variance) stays constant. However, drawing more samples from $Z$ enables us to estimate the mean of the distribution more accurately.} Consequently, we obtain the error estimate
\begin{equation}\label{eq:error_estimate_general}
\hat\sigma(\hat\xi_{XY}) = \frac{\hat\sigma((X-\langle X\rangle)(Y-\langle Y\rangle))}{\sqrt{N}}  \;\textrm{.}
\end{equation}
Here we assume that $N$ is large enough such that the likelihood function of the mean $\langle(X-\langle X\rangle)(Y-\langle Y\rangle)\rangle$ is approximately Gaussian and we are allowed to take the square-root of the variance in order to obtain a standard deviation ``$\sigma$''.

\subsection{Angular-momentum orientation}
\label{sect:autocorr_intro_L}

Our aim is to estimate the scalar two-point autocorrelation function of angular-momentum orientations, $\hat\xi_\textrm{LL}(r)$. Here, we assume spherical symmetry such that $\hat\xi(\vec r)=\hat\xi(r)$. This is a first-order approximation because the spatial distribution of galaxies in the universe is not isotropic on short scales (``Cosmic Web''). Usually, the following estimator is employed \citep[e.g.][]{Pen2000,Lee2010},
\begin{displaymath}
\hat\xi_\textrm{LL}(r) = \langle p_a p_a^\prime |\vec L_a\cdot\vec L_a^\prime|^2\rangle + \langle p_a p_b^\prime |\vec L_a\cdot\vec L_b^\prime|^2\rangle
\end{displaymath}
\begin{equation}\label{eq:estimator_LL}
+ \langle p_b p_a^\prime |\vec L_b\cdot\vec L_a^\prime|^2\rangle + \langle p_b p_b^\prime |\vec L_b\cdot\vec L_b^\prime|^2\rangle - \frac{1}{3}  \;\textrm{,}
\end{equation}
where primes indicate the second galaxy in the pair and subscripts $a, b$ denote the two possible orientations of the disc's front edge with probabilities $p_a$ and $p_b$. If the front edge is not estimated, the default values are $p_a=p_b=\frac{1}{2}$. Introducing the abbreviation $Z=p_a p_a^\prime(|\vec L_a\cdot\vec L_a^\prime|^2 + |\vec L_a\cdot\vec L_b^\prime|^2 + |\vec L_b\cdot\vec L_a^\prime|^2 + |\vec L_b\cdot\vec L_b^\prime|^2)$, an error estimate of $\hat\xi_\textrm{LL}(r)$ is
\begin{equation}\label{eq:estimator_LL_error}
\hat\sigma(\hat\xi_\textrm{LL}) = \frac{\hat\sigma(Z)}{\sqrt{N}} \;\textrm{,}
\end{equation}
where $N$ denotes the number of galaxy pairs in the relevant distance bin.

\subsection{Handedness}
\label{sect:autocorr_intro_H}

We also want to estimate the two-point autocorrelation function of handedness $\hat\xi_\textrm{HH}(r)$. Again assuming spherical symmetry, a general estimator is given by,
\begin{equation}\label{eq:estimator_HH_general}
\hat\xi_\textrm{HH}(r) = \langle h\,h^\prime\rangle \;\textrm{,}
\end{equation}
where we have defined the handedness
\begin{equation}
h = p_\textrm{Z} - p_\textrm{S} \;\textrm{.}
\end{equation}
As explained in Sect.~\ref{sect:about_galaxy_zoo}, the mean handedness is zero in the whole sample, i.e., $\langle h\rangle=\langle h^\prime\rangle=0$. Handedness alignments cannot change this in individual distance bins if the number of galaxy pairs is large enough. In every distance bin, let $n_+$ denote the number of galaxy pairs with $h\,h^\prime = +1$ and $n_-$ the number of galaxy pairs with $h\,h^\prime = -1$. We can then rewrite Eq.~(\ref{eq:estimator_HH_general}) to read
\begin{equation}\label{eq:estimator_HH}
\hat\xi_\textrm{HH}(r) = \frac{n_+ - n_-}{n_+ + n_-} = f_+ - f_- = 2f_+ - 1 \;\textrm{,}
\end{equation}
where $f_\pm = n_\pm/(n_+ + n_-)$ denotes the fraction of galaxy pairs with positive or negative handedness products, respectively. An error estimate of $\hat\xi_\textrm{HH}(r)$ is obtained from the fact that counting positive handedness products is a Bernoulli trial, i.e., $n_\pm$ are subject to the binomial distribution while $f_\pm$ are subject to the beta distribution \citep[e.g.][]{Cameron2011}.

\section{The impact of errors}
\label{sect:impact_errors}

This section is dedicated to a detailed investigation of the impact of various error sources on autocorrelation estimates of handedness and angular-momentum-orientation vectors, respectively. As key results, we finally provide marginal estimates of these autocorrelation functions which take into account all relevant error sources. Like \citet{Lee2010}, we employ isophotal ellipticity estimates as far as angular-momentum-orientation vectors are concerned. However, our methodological discussion of error propagation is also relevant in a wider context, e.g., concerning correlation functions for investigations of baryonic accoustic oscillations.

\subsection{Conditional vs.\ marginal errors}

Previous estimates \citep[e.g.][]{Slosar2009,Lee2010} employ certain input parameters such as redshift estimates using only their maximum-likelihood values, without propagating the errors of these values. Hence, these estimates are conditional instead of marginal estimates.\footnote{\citet{Slosar2009} derived pseudo-marginal estimates of the handedness autocorrelation function. Although they marginalised their likelihoods, they used conditional input data.} Consequently, we now need to explain the conceptual difference between conditional and marginal errors.

For the sake of simplicity, let us consider fitting data $D$ with Gaussian noise using a model with two \textit{linear} parameters $\theta_1$ and $\theta_2$. In this case, the likelihood function $\mathcal L(D|\theta_1,\theta_2)$ is a bivariate Gaussian also in the linear parameters and its covariance matrix
\begin{equation}
\bold\Sigma = \left(\begin{array}{cc}
\sigma_1^2 & \rho_{12}\sigma_1\sigma_2 \\
\rho_{12}\sigma_1\sigma_2 & \sigma_2^2
\end{array}\right)
\end{equation}
can be found by a Fisher analysis \citep[e.g.][]{Heavens2009}. Here, $\sigma_1^2$ and $\sigma_2^2$ denote the variances of $\theta_1$ and $\theta_2$, whereas $-1\leq\rho_{12}\leq 1$ is the correlation coefficient. $\sigma_1$ is the standard deviation of this Gaussian if sliced at the mean value of $\theta_2$. Therefore, $\sigma_1$ is the \textit{conditional} error of $\theta_1$, ``conditional'' because it depends on where the Gaussian has been sliced, i.e., the mean value of $\theta_2$. Conversely, the \textit{marginal} error of $\theta_1$ is independent of the value of $\theta_2$. This marginal error is obtained by projecting the bivariate Gaussian onto the $\theta_1$-axis, instead of slicing it. Marginal errors are never smaller than conditional errors. Consequently, the conditional error $\sigma_1$ underestimates the true error on $\theta_1$, such that, e.g., statistical significance is overestimated.

\subsection{Uncertainties in classifications}
\label{sect:uncertainties_classification}

The morphological classifications of Galaxy Zoo and \citet{HuertasCompany2011} are probabilistic, i.e., every object is assigned a probability to belong to either of the possible morphological types. This is in contrast to non-probabilistic -- ``hard'' -- assignments, where every object is clearly assigned to a certain type. Hard assignments are easier to carry out and interpret, wherefore many astronomers have a natural affinity to this approach. Unfortunately, galaxy morphologies \textit{cannot} be clearly assigned to morphological types in general -- apart from singular prototypical examples of very obvious morphology. The bulk of galaxies has uncertain morphologies, i.e., the morphological types are overlapping such that hard classification schemes are biased \citep{Andrae2010a}. For instances, a galaxy with $p_\textrm{Z} = 0.8$ still has a 20\% chance not to be a Z-spiral -- or a disc galaxy at all.\footnote{The Galaxy Zoo probabilities may exhibit minor biases due to people voting incorrectly out of confusion or malice. However, \citet{Lintott2008} weighted the users depending on how their votes agreed with the majority. Moreover, on average, every galaxy has received 39 votes \citep{Land2008} such that the impact of deliberate misclassification should give rise to a minor bias only. Certainly, that effect is much smaller than the bias we would catch up, if we cut the classification probabilities. In fact, it is very hard to do worse than a discontinuous hard cut. Any reasonable continuous transition between two classes is virtually guaranteed to be a better approximation to reality than a hard cut which corresponds to a discontinuous step in such a two-class transition.} Discarding the classification uncertainty by introducing a hard cut pretends that the data is more accurate than it actually is. This inevitably leads us to underestimate the errors, thereby compromising estimates of statistical significance.

In fact, \citet{Slosar2009} turned the probabilistic assignments of Galaxy Zoo into hard assignments by introducing a hard cut: For the clean sample, every galaxy with $p_\textrm{Z}\geq 0.8$ is considered as Z-wise spiral and every galaxy with $p_\textrm{S}\geq 0.8$ is considered as S-wise spiral, while all other galaxies are discarded. Similarly, \citet{Lee2010} considers every galaxy with $p_\textrm{Scd}^\textrm{HC}> 0.5$ as an Scd galaxy. We explain in Sects.~\ref{sect:impact_class_handedness} and \ref{sect:impact_class_orientations} how to account for these classification uncertainties in estimating the correlation functions of handedness and angular-momentum orientations. There is no reason that enforces such a hard cut.

\subsection{Errors in spectroscopic redshift estimates}

\begin{figure}
\includegraphics[width=8cm]{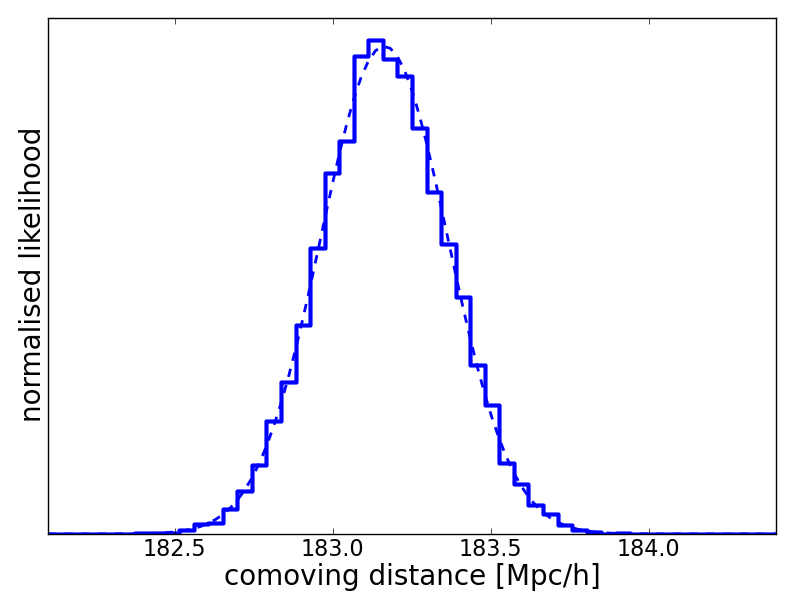}
\caption{Likelihood function of comoving distance for a galaxy with spectroscopic redshift of $z = (6.5993\pm 0.0078)\cdot 10^{-2}$. The likelihood has been estimated by drawing 10,000 Monte-Carlo samples from the error distribution of the spectroscopic redshift and is approximately Gaussian with mean $(183.16\pm 0.20)$Mpc/h.}
\label{fig:errors_in_comoving_distance}
\end{figure}

Both autocorrelation functions introduced in Sects.~\ref{sect:autocorr_intro_L} and~\ref{sect:autocorr_intro_H} require estimates of distances of galaxy pairs and these distances are uncertain due to errors in the redshift estimates. In order to assess the impact of redshift errors, we randomly select a single galaxy from our SDSS subsample and draw 10,000 Monte-Carlo samples from its redshift-error distribution. For every sampled value of redshift, we compute the comoving distance and monitor its distribution. As is evident from Fig.~\ref{fig:errors_in_comoving_distance}, the errors in the comoving distances are of the same order of magnitude as the typical distance scale of the correlations reported in the literature ($\approx 1\Mpc/h$). Consequently, these errors are important and have to be taken into account. We explain in Sect.~\ref{sect:propagating_errors} how to propagate these redshift errors by Monte-Carlo sampling.

\subsection{Errors in ellipticity estimates}
\label{sect:eps_errors_general}

Errors in ellipticity estimates used as proxies for disc inclination clearly have an impact on the estimation of the angular-momentum orientations and their correlation function. We now try to estimate these errors. We explain in Sect.~\ref{sect:propagating_errors} how to propagate ellipticity errors by Monte-Carlo sampling.

\begin{figure}
\includegraphics[width=8cm]{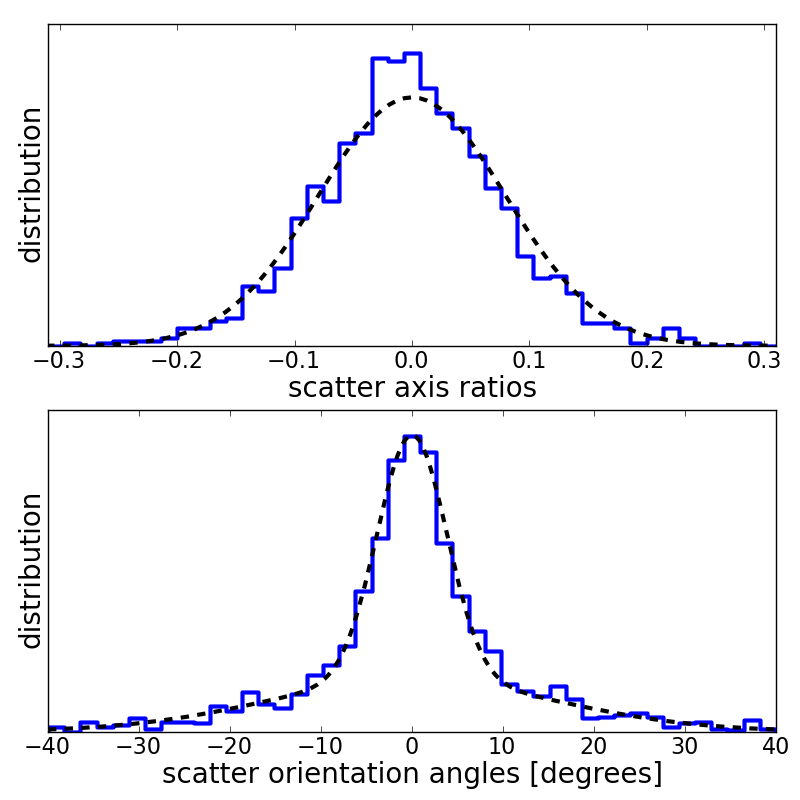}
\caption{Distributions of differences in isophotal axis ratios (top) and isophotal orientation angles (bottom) for the 1,596 rogue pairs in the catalogue of \citet{HuertasCompany2011}. The top panel is well approximated by a Gaussian with mean zero and standard deviation of $\approx 0.0795$ (dashed line), which yields an error estimate of $\hat\sigma(q_\textrm{iso})=\frac{0.0795}{\sqrt{2}}\approx 0.0562$. The distribution of differences in orientation angles (bottom panel) is not Gaussian, but described by the ad-hoc model of Eq.~(\ref{eq:likelihood_delta_theta}) (dashed line) based on Eq.~(\ref{eq:error_estimate_iso_orientation}) with manually adjusted parameters $\hat\alpha\approx 0.73$, $\hat\sigma_1\approx 2.7^\circ$ and $\hat\sigma_2\approx 15.0^\circ$.}
\label{fig:estimation_isophotal_ellipticity_errors}
\end{figure}

First, considering the isophotal ellipticities used by \citet{Lee2010}, the SDSS database unfortunately does not offer error estimates.\footnote{In fact, the table \textsc{galaxy} contains columns for the errors of the isophotal ellipticities. However, for the relevant objects these columns are only filled with invalid default values.} Consequently, employing isophotal ellipticites, the SDSS database strictly does not enable us to estimate a marginal autocorrelation function. In order to get a rough estimate of the errors in isophotal ellipticities, we make use of the rogue pairs in the SDSS database, i.e., multiple entries of identical galaxies. Starting out from 698,420 galaxies in the classification table provided by \citet{HuertasCompany2011}, we identify rogue pairs as galaxy pairs whose angular separation is less than 0.4 arcsec, which roughly corresponds to one pixel size.\footnote{Here, we assume that for multiple entries the whole galaxy is used for parameter estimation and not only a shredded part of the galaxy.} We find 1,596 such pairs. We then monitor the difference in axis ratios and orientation angles of every pair. The resulting distributions are shown in Fig.~\ref{fig:estimation_isophotal_ellipticity_errors}. As rough error estimate for the isophotal axis ratio, we obtain a standard deviation of
\begin{equation}\label{eq:error_estimate_iso_axis_ratio}
\hat\sigma(q_\textrm{iso})\approx 0.0562 \;\textrm{,}
\end{equation}
when fixing the mean to zero. The distribution of differences in orientation angles is not Gaussian but has more prominent wings. We therefore model the likelihood function of orientation angles with mean angle $\theta_0$ as a mixture of two Gaussians of different width,
\begin{equation}\label{eq:error_estimate_iso_orientation}
\mathcal L(\theta|\theta_0,\sigma_1,\sigma_2,\alpha) = \alpha N(\theta|\theta_0,\sigma_1) + (1-\alpha) N(\theta|\theta_0,\sigma_2) \;\textrm{.}
\end{equation}
The bottom panel of Fig.~\ref{fig:estimation_isophotal_ellipticity_errors} displays the distribution of differences of two values drawn from Eq.~(\ref{eq:error_estimate_iso_orientation}), whose likelihood is obtained by convolving $\mathcal L(\theta|\theta_0,\sigma_1,\sigma_2,\alpha)$ with itself. The resulting likelihood function then reads
\begin{displaymath}
\mathcal L(\Delta\theta|\sigma_1,\sigma_2,\alpha) = \alpha^2 N(\Delta\theta|0,\sqrt{2}\sigma_1) 
\end{displaymath}
\begin{equation}\label{eq:likelihood_delta_theta}
+ 2\alpha(1-\alpha) N(\Delta\theta|0,\sqrt{\sigma_1^2+\sigma_2^2}) + (1-\alpha)^2 N(\Delta\theta|0,\sqrt{2}\sigma_2) \;\textrm{.}
\end{equation}
Manually adjusting the model parameters of Eq.~(\ref{eq:error_estimate_iso_orientation}), we obtain a rough error estimate for the isophotal orientation angle with parameters $\hat\alpha\approx 0.73$, $\hat\sigma_1\approx 2.7^\circ$ and $\hat\sigma_2\approx 15.0^\circ$. If we required an angular separation of 0, i.e., identical coordinates, we would still end up with 17 pairs exhibiting similar scatter in both parameters.

Second, the correction for intrinsic axial ratios of Scd galaxies is subject to uncertainties, too. Wherever we neglect ellipticity errors, we also neglect errors in intrinsic axial ratios and simply adopt $p=0.1$. Conversely, if we take into account ellipticity errors, we will automatically also take into account errors in the intrinsic axial ratio. In this case, we assume that $p$ is drawn from a uniform distribution over the interval $[0.083,0.145]$ (see Sect.~\ref{sect:from_q_to_L}).

\subsection{Propagating errors numerically}
\label{sect:propagating_errors}

We now explain how to incorporate errors in redshift estimates and ellipticity estimates. The crucial problem is that both errors cannot be propagated analytically.

We propagate the measurement errors of spectroscopic redshift and ellipticity by drawing 1,000 Monte-Carlo realisations from the error distributions of both parameters and averaging the results over all Monte-Carlo realisations.\footnote{Analysing multiple Monte-Carlo realisations is of course computationally expensive. However, this task is still easily executed on a standard computer.} A value for the intrinsic axial ratio is drawn from the uniform interval $[0.083,0.145]$ once for every Monte-Carlo realisation, i.e., in each realisation all galaxies have the same correction for intrinsic axial ratio. This Monte-Carlo sampling is in fact a marginalisation over the errors of both observables, spectroscopic redshift and ellipticity. Typically, both error sources are not taken into account \citep[e.g.][]{Slosar2009,Lee2010}, which yields correlation estimates with \textit{conditional} errors -- conditional because they assume, e.g., the observed redshifts were the true ones.

A final remark concerning the correlation estimate: We monitor the distribution of the correlation values $\hat\xi$ resulting from the 1,000 Monte-Carlo realisations. However, a fundamental difference to Eq.~(\ref{eq:error_estimate_general}) is that now $\hat\xi$ itself is a random variate. Consequently, we are now interested in the variance of $\hat\xi$ but not in the variance of the mean of $\hat\xi$. The difference is a factor of 1,000 in the variances. It is obvious that this approach is correct, since otherwise we could make the resulting errors arbitrarily small by increasing the number of Monte-Carlo realisations.

\subsection{Negligible error sources}

There are further sources of errors which could be taken into account but are not relevant in our case.

For instances, uncertainties in the cosmological parameters have an impact on the comoving distances. In our case, this is irrelevant because all galaxies are affected the same way. However, if in a different context the task is to use marginal autocorrelation functions in order to do cosmological inference, it may be mandatory to also incorporate uncertainties of cosmological parameters into the Monte-Carlo sampling described in Sect.~\ref{sect:propagating_errors}. We experienced that this increases the error in comoving distances by approximately a factor of two.

Another negligible error source is the position estimate of a galaxy in equatorial coordinates, which is obviously much smaller than, e.g., any redshift error. Given the pixel size of $\approx 0.4$~arcsec of SDSS, at a redshift of $z=0.066$ and comoving distance of $d=183$Mpc/h one pixel misestimation corresponds to a transversial error of 0.35kpc/h. This is several orders of magnitude below the theoretically expected correlation length of roughly 1Mpc/h \citep{Schaefer2011}.

\begin{figure*}
\includegraphics[width=15cm]{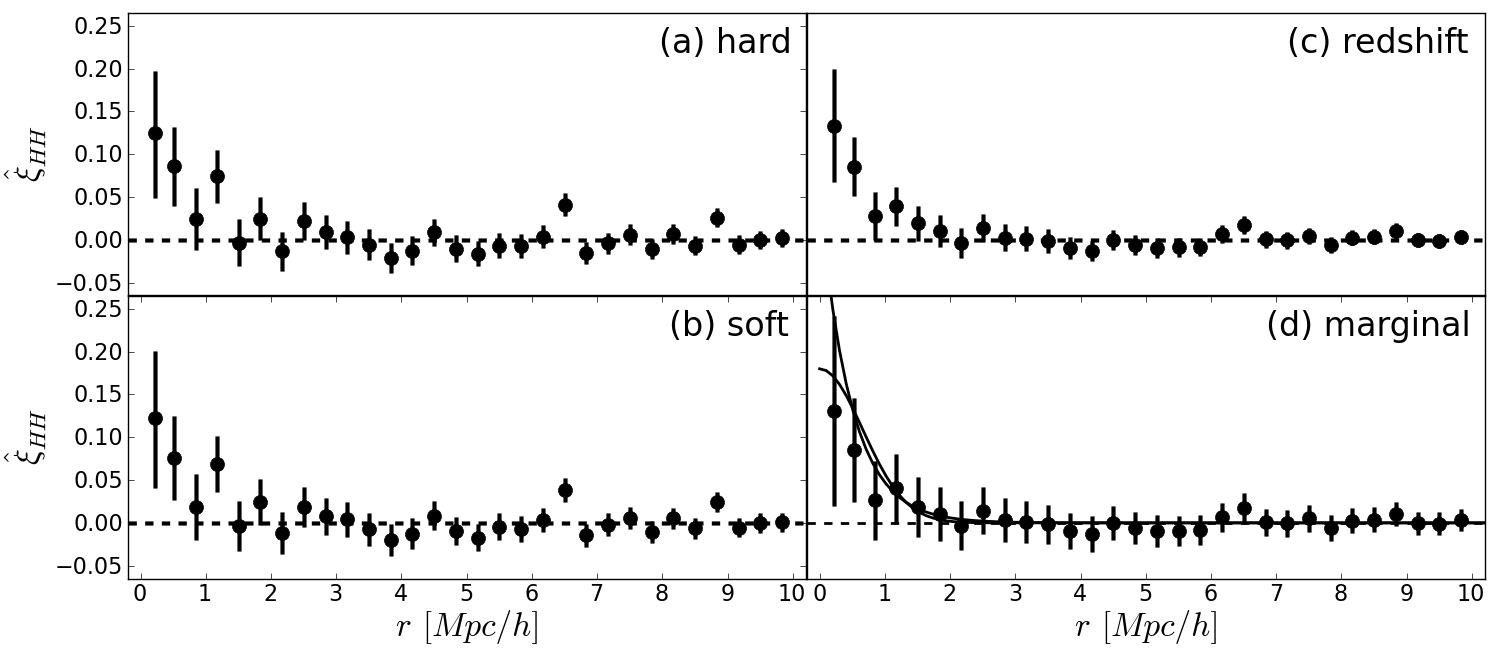}
\caption{Impact of various errors on estimates of the handedness autocorrelation function. Panel (a): Hard estimate neglecting classification uncertainties and redshift errors, taking into account only number statistics. Panel (b): Soft estimate accounting for classification uncertainties and number statistics, neglecting redshift errors. Panel (c): Estimate accounting for redshift errors and number statistics, ignoring classification uncertainties. Panel (d): Marginal estimate taking into account classification uncertainties, redshift errors and number statistics. Furthermore, we show autocorrelation estimates parametrised as exponential and Gaussian according to Fig.~\ref{fig:handedness_correlations_likelihood_manifolds}.}
\label{fig:impact_errors_on_handedness_autocorrelation}
\end{figure*}

\subsection{Impact on autocorrelation of handedness}
\label{sect:impact_class_handedness}

In this section, we discern the impact of various error sources on estimates of the handedness autocorrelation function, namely classification uncertainties, redshift errors and number statistics. Our ultimate goal is a marginal estimate of the handedness autocorrelation function, where all errors have been marginalised out.

First, we use the hard estimator from \citet{Slosar2009}, which does not account for uncertainties in classification and redshift. Panel (a) of Fig.~\ref{fig:impact_errors_on_handedness_autocorrelation} shows our estimate of this autocorrelation function for the Galaxy Zoo sample. Qualitatively, our results agree with the results of \citet{Slosar2009}. We observe positive correlations, i.e., an alignment of handedness, on short distances, too. Although there are minor differences which might arise from the slightly different data sets used, the general agreement validates our method.

Second, we take into account uncertainties in the handedness classifications, but still ignore redshift errors. In every distance bin, we compute the handedness products
\begin{equation}\label{eq:handedness_product}
h\,h^\prime = (p_\textrm{Z}-p_\textrm{S})(p_\textrm{Z}^\prime-p_\textrm{S}^\prime)
= p_\textrm{Z}p_\textrm{Z}^\prime + p_\textrm{S}p_\textrm{S}^\prime - p_\textrm{Z}p_\textrm{S}^\prime - p_\textrm{S}p_\textrm{Z}^\prime \;\textrm{,}
\end{equation}
which can now take any value in the interval $[-1,1]$. The correlation estimator of Eq.~(\ref{eq:estimator_HH}) is unchanged. However, $n_\pm$ now are \textit{not} the number of pairs where $h\,h^\prime=\pm 1$, but are rather defined by
\begin{equation}
n_+ = \sum_\textrm{pairs}(p_\textrm{Z}p_\textrm{Z}^\prime + p_\textrm{S}p_\textrm{S}^\prime) \quad\textrm{and}\quad
n_- = \sum_\textrm{pairs}(p_\textrm{Z}p_\textrm{S}^\prime + p_\textrm{S}p_\textrm{Z}^\prime) \;\textrm{.}
\end{equation}
Note that if $N$ denotes the number of galaxy pairs in a given distance bin, then $n_++n_-\leq N$. Consequently, this reduces the ``effective'' number of galaxy pairs in a given distance bin because the contribution of every galaxy pair is downweighted by the probability that either galaxy is not a spiral with handedness. Furthermore, reducing the effective number of galaxy pairs also increases the error of the correlation estimate through the beta distribution \citep[e.g.][]{Cameron2011}. Results of this estimator are shown in panel (b) of Fig.~\ref{fig:impact_errors_on_handedness_autocorrelation}. As expected, the errorbars are indeed slightly larger.

Third, we account for redshift errors but ignore classification uncertainties. As described in Sect.~\ref{sect:propagating_errors}, we draw 1,000 Monte-Carlo realisation from the error distributions of the spectroscopic redshift estimates and average over all realisations. Panel (c) of Fig.~\ref{fig:impact_errors_on_handedness_autocorrelation} shows the resulting estimate of the handedness autocorrelation. In comparison to panels (a) and (b), the autocorrelation function now looks remarkably smooth. Errors in redshift cause uncertainties in the distances, i.e., galaxy pairs end up in different distance bins in different realisations. Consequently, a likely explanation for all the substructures in panels (a) and (b) is that they are noise features that have been enhanced by binning.

Finally, panel (d) shows the marginal autocorrelation function, which takes into account all important sources of uncertainty. The errorbars are so large that apparently no statistically significant positive correlation of handedness can be detected. However, we have to refine this question in the next section, as we should not attempt to assess statistical significance from binned data.

\subsection{Parameter estimation}
\label{sect:para_est_handedness}

Figure~\ref{fig:impact_errors_on_handedness_autocorrelation} shows binned versions of the estimated correlation function. This is acceptable as long as we only study the dependence of the errorbars on the different error sources. However, in order to assess the statistical significance of positive autocorrelations in the final marginal estimate, we should try to avoid the ambiguities introduced by binning. For this purpose, we employ the likelihood function of the data $D$ introduced by \citet{Slosar2009},
\begin{equation}\label{eq:handedness_likelihood}
\mathcal L[D|\xi(r)] = \prod_{\textrm{pairs }p}\left(\frac{1 + d_p\,\xi(r_p)}{2}\right) \,\textrm{,}
\end{equation}
where $r_p$ is the distance between the two galaxies of the $p$-th pair. The coefficient $d_p$ is the handedness product of both galaxies. As \citet{Slosar2009} used hard cuts of the Galaxy Zoo classifications, $d_p=\pm 1$ in their case. We modify this by equating $d_p$ with Eq.~(\ref{eq:handedness_product}) such that now $-1\leq d_p\leq+1$ and galaxy pairs are weighted by the probability that both of them are spirals.

In order to assess the statistical significance of potential positive autocorrelations in spiral-arm handedness, we follow \citet{Slosar2009} in using the Bayes factor,
\begin{equation}\label{eq:bayes_factor}
\frac{\prob(D|\mathcal M_+)}{\prob(D|\mathcal M_0)} = \frac{\int\prob(D|\theta_+,\mathcal M_+)\,\prob(\theta_+|\mathcal M_+)d\theta_+}{\int\prob(D|\theta_0,\mathcal M_0)\,\prob(\theta_0|\mathcal M_0)d\theta_0} \,\textrm{.}
\end{equation}
Here, $\prob(D|\mathcal M_n)$ denotes the likelihood of the data $D$ given the model $\mathcal M_n$, \textit{irrespective} of what the parameter values $\theta_n$ of model $\mathcal M_n$ are. Conversely, $\prob(D|\theta_n,\mathcal M_n)$ denotes the likelihood of the data given the model $\mathcal M_n$ and certain parameter values $\theta_n$, while $\prob(\theta_n|\mathcal M_n)$ denotes the prior probability of the parameter values $\theta_n$ of model $\mathcal M_n$.\footnote{If we assume that both models, $\mathcal M_+$ and $\mathcal M_0$, are equally likely a-priori, i.e., if we have no a-priori preference, then the Bayes factor is identical to the ratio of model posteriors, which quantify the probability of the model given the data.}

\begin{figure}
\includegraphics[width=8cm]{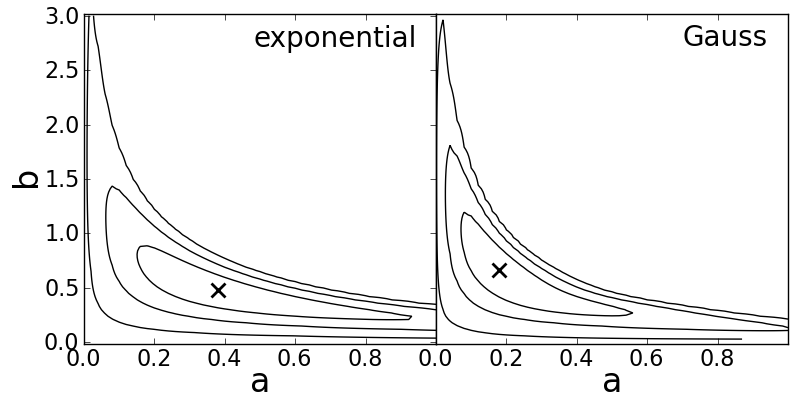}
\caption{Likelihood contours constraining $a$-$b$ plane of exponential (left) and Gaussian (right) handedness-autocorrelation functions. The likelihood maximum is indicated by a cross and the contours enclose 68.3\%, 95.5\%, and 99.7\% confidence. The maximum of the exponential model occurs at $a=0.38$ and $b=0.48\Mpc/h$. The maximum of the Gaussian model occurs at $a=0.18$ and $b=0.66\Mpc/h$.}
\label{fig:handedness_correlations_likelihood_manifolds}
\end{figure}

In our case, the model $\mathcal M_0$ describes the null hypothesis that no autocorrelation exists, i.e., $\xi(r) = 0$. This model has no free parameters, such that we can directly evaluate $\prob(D|\mathcal M_0)$ via Eq.~(\ref{eq:handedness_likelihood}). Conversely, the model $\mathcal M_+$ is supposed to describe positive autocorrelations. Here, we have to make a choice how we parametrise such positive autocorrelations. Like \citet{Slosar2009}, we then employ two parametrisations, an exponential and a Gaussian,
\begin{equation}\label{eq:handedness_likelihood_parametrisations}
\xi_\textrm{exp}(r) = a\,e^{-r/b} \quad\textrm{and}\quad \xi_\textrm{Gauss}(r) = a\,e^{-r^2/2b^2} \,\textrm{,}
\end{equation}
with model amplitudes $a$ and model correlation lengths $b$. For both models, we use flat and \textit{normalised} priors within the intervals $a\in(0,1]$ and $b\in(0,3]$. In contrast to \citet{Slosar2009}, we also exclude $a=0$ in order to ensure that $\mathcal M_+$ and $\mathcal M_0$ are indeed mutually exclusive. As both parametrisations introduced in Eq.~(\ref{eq:handedness_likelihood_parametrisations}) have two free parameters, $a$ and $b$, we cannot evaluate $\mathcal M_+$ directly. Rather, we compute the likelihood manifolds of $a$ and $b$ for both models using a brute-force grid. Figure~\ref{fig:handedness_correlations_likelihood_manifolds} shows the resulting likelihood manifolds averaged over the 1,000 noise realisations drawn from the redshift-error distribution. Our results look very similar to those shown in Fig.~4 of \citet{Slosar2009} and our most likely values agree nicely with their values. The best-fit estimates for both models are also shown in panel (d) of Fig.~\ref{fig:impact_errors_on_handedness_autocorrelation}. Given the brute-force likelihood grid $\mathcal L_{ij}=\mathcal L(a_i,b_j)$, the marginalisation integral in Eq.~(\ref{eq:bayes_factor}) can be approximated by a Riemann sum,
\begin{displaymath}
\int_0^1 da\int_0^3db\,\prob(D|a,b,\mathcal M_+)\,\prob(a,b|\mathcal M_+)
\end{displaymath}
\begin{equation}
\approx \sum_{i,j}\,\mathcal L(a_i,b_j)\Delta a_i\Delta b_j\frac{1}{3} \,\textrm{,}
\end{equation}
where $\prob(a,b|\mathcal M_+)=\frac{1}{3}$ is the normalised flat prior on the interval $a\in(0,1]$ and $b\in(0,3]$, while $\Delta a_i=\Delta a$ and $\Delta b_j=\Delta b$ denote the equidistant stepsizes of the brute-force likelihood grids shown in Fig.~\ref{fig:handedness_correlations_likelihood_manifolds}. This results in Bayes factors of 27.9 for the exponential model\footnote{This means that it is 27.9 times more likely that the data has been drawn from an exponential whose amplitude is somewhere in the range $(0,1]$ and scale radius is somewhere in $(0,3]\Mpc/h$ than that the data has been drawn from a zero correlation function.} and 13.1 for the Gaussian model, respectively. These values can be interpreted as strong \textit{but not yet decisive} evidence in favour of positive autocorrelations. Decisive evidence requires Bayes factors larger than 100 \citep[e.g.][]{Kass1993}.

\subsection{Impact on autocorrelation of angular-momentum orientation}
\label{sect:impact_class_orientations}

Now, we discern the impact of various error sources on estimates of the autocorrelation function of angular-momentum orientation vectors. Again, our ultimate goal is a marginal estimate of the handedness autocorrelation function.

\begin{figure*}
\includegraphics[width=15cm]{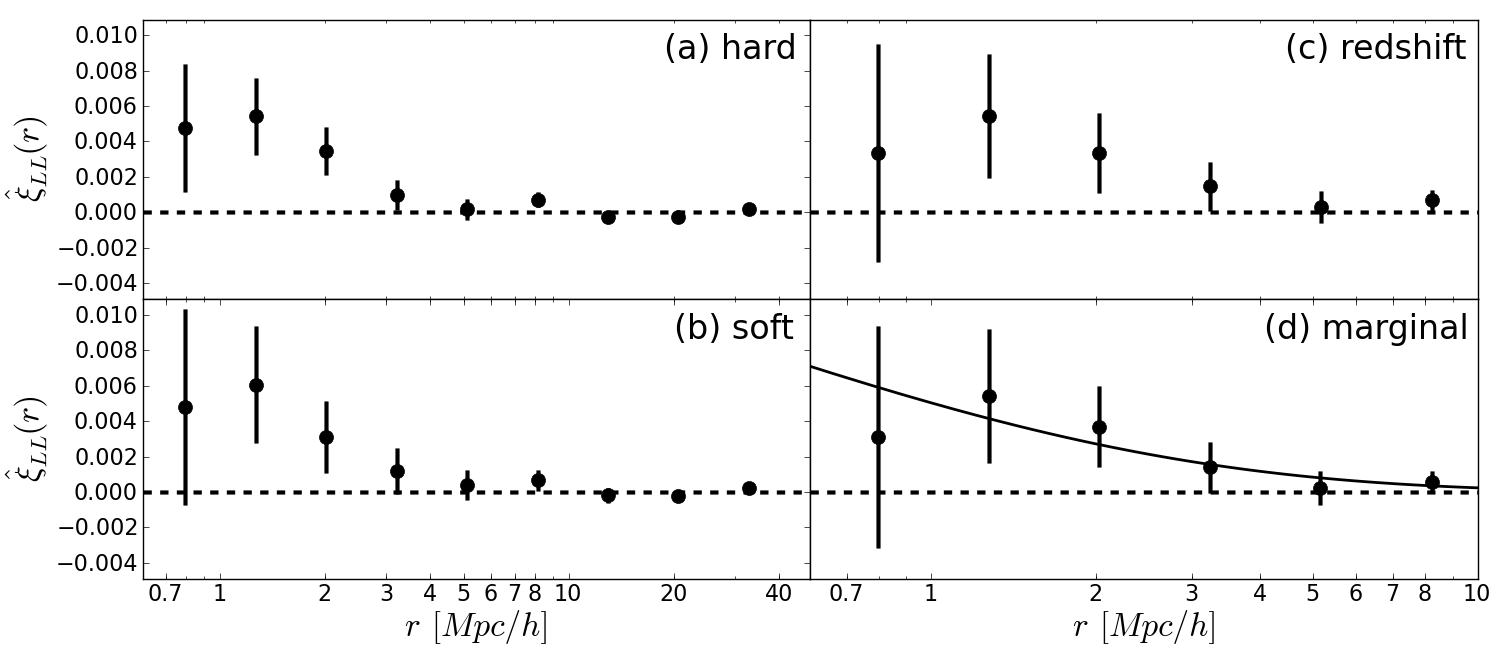}
\caption{Impact of various errors on estimates of autocorrelation function of angular-momentum-orientation vectors. Panel (a): Hard estimator, neglecting classification uncertainties and errors in redshift and ellipticity estimates, taking into account only number statistics. Panel (b): Soft estimator, accounting for classification uncertainties and number statistics, neglecting errors in redshift and ellipticity estimates. Panel (c): Conditional estimate accounting for redshift errors, classification uncertainties and number statistics, neglecting errors in ellipticity estimates. Panel (d): Marginal estimate taking into account classification uncertainties, number statistics, and errors in redshift and ellipticity estimates. The solid line represents the fit given by Eq.~(\ref{eq:LL_correlation_fit}).}
\label{fig:impact_errors_on_LL_autocorrelation}
\end{figure*}

First, we try to reproduce the estimate of angular-momentum-orientation autocorrelation of \citet{Lee2010}. The only difference is that we have removed 20 objects from the galaxy sample in order to eliminate rogue pairs. Panel (a) of Fig.~\ref{fig:impact_errors_on_LL_autocorrelation} shows our resulting estimate of the autocorrelation via Eq.~(\ref{eq:estimator_LL}). Our result is identical to the one of \citet{Lee2010}. This implies that, first, our method is working correctly, and, second, that a few rogue pairs have negligible impact on the results of \citet{Lee2010}.

Second, we study the impact of uncertainties of morphological classification. Formally, the estimator defined in Eq.~(\ref{eq:estimator_LL}) does not change, only the effective number of galaxy pairs in all redshift bins is reduced. Picking out a single of the four terms in Eq.~(\ref{eq:estimator_LL}), we change the definition
\begin{equation}
\langle p_q p_a^\prime |\vec L_a\cdot\vec L_a^\prime|^2\rangle = \frac{\sum_\textrm{pairs}p_\textrm{Scd}^\textrm{HC}{p_\textrm{Scd}^\textrm{HC}}^\prime p_q p_a^\prime |\vec L_a\cdot\vec L_a^\prime|^2}{\sum_\textrm{pairs}p_\textrm{Scd}^\textrm{HC}{p_\textrm{Scd}^\textrm{HC}}^\prime} \;\textrm{.}
\end{equation}
This weights the contribution of every pair by the probability $p_\textrm{Scd}^\textrm{HC}{p_\textrm{Scd}^\textrm{HC}}^\prime$ that \textit{both} galaxies are Scd galaxies. Furthermore, the number $N$ of pairs in the distance bin are replaced by the sum of weights $\sum_\textrm{pairs}p_\textrm{Scd}^\textrm{HC}{p_\textrm{Scd}^\textrm{HC}}^\prime\leq N$. Obviously, this weighting also affects the error estimate of Eq.~(\ref{eq:estimator_LL_error}). Panel (b) of Fig.~\ref{fig:impact_errors_on_LL_autocorrelation} shows the probabilistic correlation estimate. Evidently, the hard estimator used by \citet{Lee2010} substantially underestimates the errors, thereby overestimating the actual statistical significance. As class probabilities were cut at $p_\textrm{Scd}^\textrm{HC}>0.5$, classification uncertainties have a larger impact than in the case of handedness where the cut of handedness probabilities was at 0.8.

Third, we incorporate errors in spectroscopic redshift by drawing 1,000 Monte-Carlo realisations from the redshift's error distribution. The resulting conditional estimate, now out to $10\Mpc/h$, is shown in panel (c) of Fig.~\ref{fig:impact_errors_on_LL_autocorrelation}. Qualitatively, the impact of redshift errors on the correlation estimate of angular-momentum-orientation vectors is not as severe as in the case of handedness (cf.\ marginal estimate of Fig.~\ref{fig:impact_errors_on_handedness_autocorrelation}). Note, the binsize in Fig.~\ref{fig:impact_errors_on_LL_autocorrelation} is much larger than in Fig.~\ref{fig:impact_errors_on_handedness_autocorrelation} because here we are studying a smaller sample with fewer galaxy pairs. Nonetheless, the estimated errors have indeed increased, which is particularly obvious for the first distance bin. As the binning is logarithmic in distance, this is not surprising because the first distance bin has the smallest binsize and is thereby strongest affected by redshift errors ``smearing out'' galaxy pairs along the horizontal axis.\footnote{We do not expect distance errors of the order of 0.2Mpc/h to have a large impact on a distance bin of 1Mpc/h binsize.}

Finally, we also take into account errors in ellipticity estimates. As mentioned in Sect.~\ref{sect:eps_errors_general}, the SDSS database actually does not provide error estimates for the isophotal ellipticities. Hence, we need to proceed using the rough error estimates of Eqs.~(\ref{eq:error_estimate_iso_axis_ratio}) and~(\ref{eq:error_estimate_iso_orientation}) as well as the uniform error in intrinsic axis ratios. This enables us to estimate a marginal autocorrelation function which is shown in panel (d) of Fig.~\ref{fig:impact_errors_on_LL_autocorrelation}. In comparison to panel (c), there is only a minor increase in the errorbars. However, we would not put too much faith into the marginal estimate because the error estimate of ellipticities is rather coarse. Nevertheless, comparing to panel (a), the marginal estimate differs substantially from a conditional estimate and there are no statistically significant autocorrelations.

\subsection{Constraining theoretical parameters}
\label{sect:LL_fitting}

The autocorrelation of angular-momentum orientations can be used to estimate free parameters in the tidal-torque theory \citep[e.g.][]{Lee2008}. Let $\xi(r,R)$ denote the two-point correlation function of the density field, smoothed over scale $R$. In this case, one can derive a model prediction for the linear regime \citep[e.g.][]{Pen2000}
\begin{equation}
\xi_\textrm{LL}(r) \approx \frac{a^2}{6} \frac{\xi^2(r,R)}{\xi^2(r,0)} \,\textrm{,}
\end{equation}
where $a$ is a free model parameter. For the nonlinear regime, \citet{Lee2008} derived the following model prediction
\begin{equation}
\xi_\textrm{LL}(r) \approx \frac{a_\textrm{L}^2}{6} \frac{\xi^2(r,R)}{\xi^2(r,0)} + \varepsilon_\textrm{NL}\frac{\xi(r,R)}{\xi(r,0)} \,\textrm{,}
\end{equation}
where $a_\textrm{L}$ and $\varepsilon_\textrm{NL}$ are free model parameters describing the linear and nonlinear contributions. Estimating values for these model parameters is important in order to constrain the tidal-torque theory.\footnote{In fact, this is the reason why \citet{Lee2010} restricts the sample to galaxies with $z\leq 0.02$ in order to obtain a volume-limited sample. Otherwise, the density field of galaxies cannot be meaningfully defined and $\xi(r,R)$ cannot be estimated.} The impact of the additional error sources on this parameter estimation is devastating. First, the marginal estimate of $\xi_\textrm{LL}(r)$ has large errors. Second, errors in redshift estimates and morphological classification\footnote{As $\hat\xi_\textrm{LL}(r)$ has been estimated for Scd galaxies, also $\xi(r,R)$ has to be estimated for this type of galaxies.} also affect the estimation of the two-point correlation function $\xi(r,R)$. Given these considerations and the SDSS sample, we have to conclude that it is currently impossible to place decisive constraints on the theoretical parameters.

\begin{figure}
\begin{center}
\includegraphics[width=8cm]{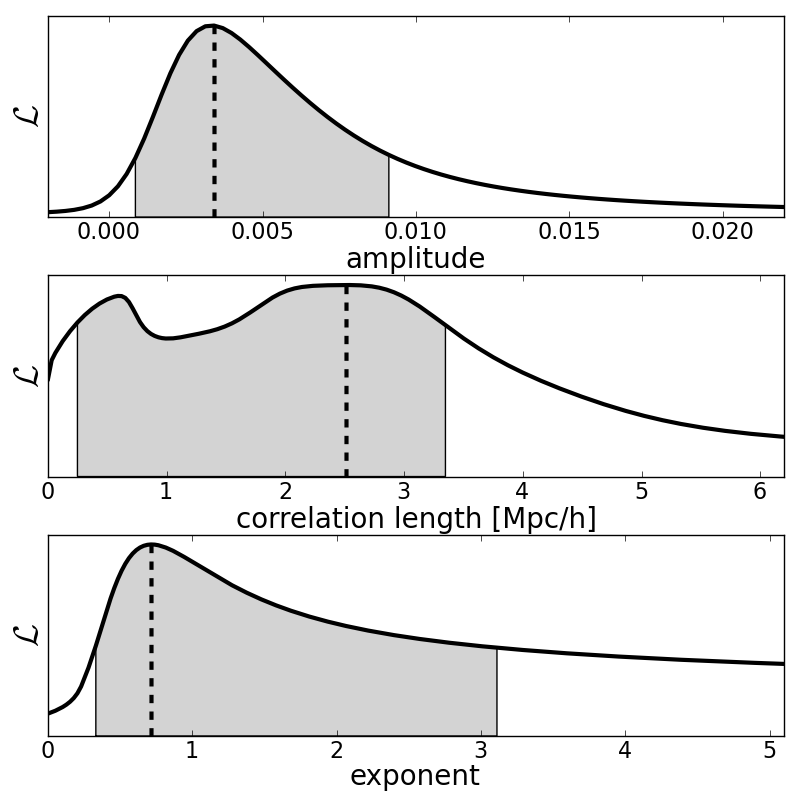}
\end{center}
\caption{Marginal likelihoods of fitting the marginal angular-momentum-orientation autocorrelation. The model given by Eq.~(\ref{eq:LL_generic_model_Schaefer}) is fitted to the binned version of the marginal autocorrelation of Fig.~\ref{fig:impact_errors_on_LL_autocorrelation}d. Top panel: Marginal likelihood of amplitude with maximum at $A=0.0034_{-0.0027}^{+0.0057}$. Centre panel: Marginal likelihood of correlation length with maximum at $R=2.5_{-2.3}^{+0.8}$. Bottom panel: Marginal likelihood of exponent with maximum at $C=0.71_{-0.38}^{+2.40}$. The asymmetric errors denote 68\% confidence intervals. The parameter estimation has been conducted on a three-dimensional brute-force grid. As the distributions of Monte-Carlo realisations in every distance bin are Gaussian in excellent approximation, the fit is done via $\chi^2$-minimisation.}
\label{fig:LL_correlation_fitting_full_with_conf_levels}
\end{figure}

The same argument applies to the generic autocorrelation model proposed by \citet{Schaefer2011},
\begin{equation}\label{eq:LL_generic_model_Schaefer}
\xi_\textrm{LL}(r) = A\exp\left[-\left(\frac{r}{R}\right)^C\right] \,\textrm{,}
\end{equation}
which contains a linear amplitude $A$ and two nonlinear model parameters $R$ and $C$ that cannot be constrained properly. Figure~\ref{fig:LL_correlation_fitting_full_with_conf_levels} demonstrates this by showing the marginal likelihoods of fitting Eq.~(\ref{eq:LL_generic_model_Schaefer}) to the binned data of panel (d) of Fig.~\ref{fig:impact_errors_on_LL_autocorrelation}.\footnote{Actually, we should estimate the correlation function from unbinned data like in Sect.~\ref{sect:para_est_handedness}. However, a meaningful likelihood function is not easily defined in this case such that we have to resort to fitting binned data. We are fully aware that binning may compromise our assessment of statistical significance.} Evidently, the (marginal) uncertainties in all model parameters are extremely large. Nevertheless, let us note that the correlation length of $1\Mpc/h$ predicted by \citet{Schaefer2011} is in agreement with our estimate. Furthermore, for later purposes, we identify the best fitting model,\footnote{Note that the maximum of the \textit{joint} likelihood does not coincide with the maxima of the \textit{marginalised} likelihoods in Fig.~\ref{fig:LL_correlation_fitting_full_with_conf_levels}.}
\begin{equation}\label{eq:LL_correlation_fit}
\xi_\textrm{LL}(r) \approx 0.026\cdot\exp\left[-\left(\frac{r}{0.34\Mpc/h}\right)^{0.46}\right] \,\textrm{.}
\end{equation}
We explicitly emphasise that we do \textit{not} claim that this were by any means a model of the true correlation function. This fit is solely meant to provide us with \textit{some} model that is \textit{compatible} with the data. Such a model is later required in order to conduct simulations. This is also the reason why we do not need to estimate errors for the fit given by Eq.~(\ref{eq:LL_correlation_fit}).

\section{Biased ellipticity estimates from second moments}
\label{sect:eps_biases_2nd_moments}

Isophotal ellipticity estimates have the disadvantage that they strongly depend on the choice of a particular isophote and therefore may suffer strongly from pixel noise. Ellipticity estimates based on the moments of the galaxy's light distribution at first glance seem to be more promising, since no isophote is required and the complete data enters the estimate. Consequently, we would expect that ellipticity estimates based on light moments are more robust against pixel noise than isophotal ellipticities which might improve autocorrelation estimates of angular-momentum-orientation vectors. However, in this section, we demonstrate that ellipticity estimates based on second moments of the light distribution are so strongly biased that they cannot be used for investigations of disc alignment. In particular, this bias would cause us to overestimate the correlation due to alignment such that, e.g., we would overestimate its impact on weak-lensing studies.

\subsection{Revealing the bias}

\begin{figure}
\includegraphics[width=8cm]{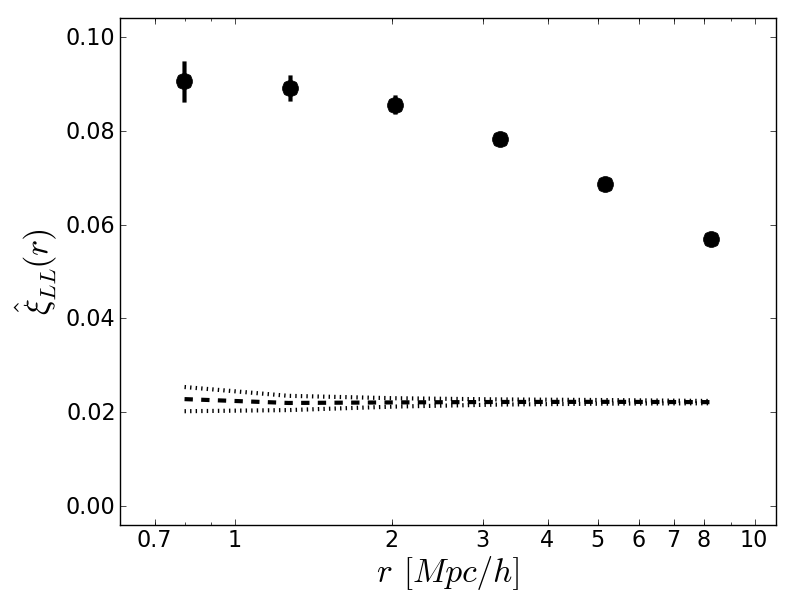}
\caption{Pseudo-marginal estimate of the angular-momentum-orientation autocorrelation for the sample of Scd galaxies taking into account uncertainties in classification, number statistics, errors in redshift and ellipticity estimates. Results have been averaged over 1,000 Monte-Carlo samples drawn from the error distribution of spectroscopic redshifts. The dots indicate mean values and the errorbars correspond to one Gaussian standard deviation. The horizontal dashed line indicates the background correlation level estimated from 100 random shufflings of the galaxy positions, a method used by Lee (2011).}
\label{fig:LL_correlation_2moments}
\end{figure}

\begin{figure}
\includegraphics[width=8cm]{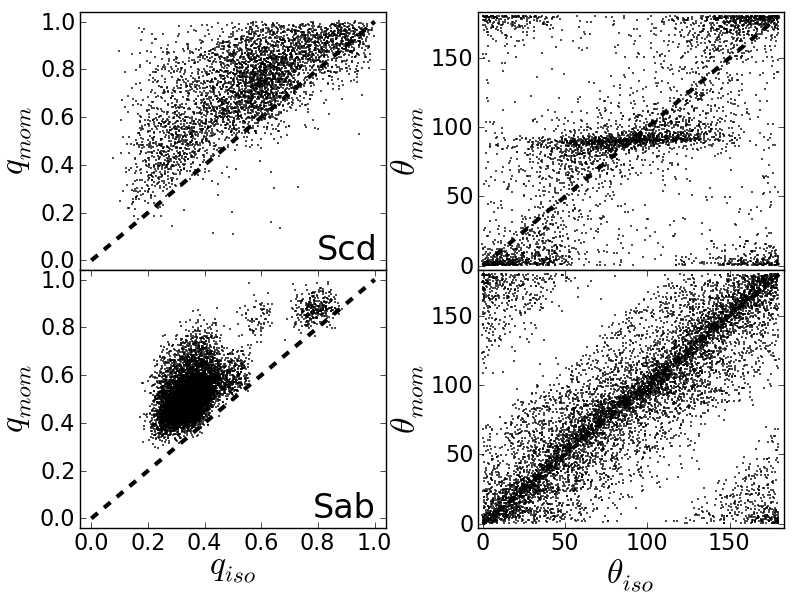}
\caption{Comparing ellipticities based on isophotes and unweighted second moments. Top panels: Axis ratios (left) and orientation angles (right) for Scd galaxies. Bottom panels show the same for Sab galaxies. Axis ratios estimated from second moments are systematically larger than those estimated from isophotal contours, i.e., second moments find the disc galaxies to be rounder. Orientation angles are unbiased. The distributions of axis ratios for Scd and Sab galaxies agrees with the results of \citet{HuertasCompany2011} (their Fig.~2).}
\label{fig:compare_isophotal2Stokes_compact}
\end{figure}

We also assess the usage of ellipticity estimates based on unweighted second moments of the galaxies' light distributions. Furthermore, SDSS offers error estimates for these parameters. Figure~\ref{fig:LL_correlation_2moments} shows the result. The most striking difference to Fig.~\ref{fig:impact_errors_on_LL_autocorrelation}d is that Fig.~\ref{fig:LL_correlation_2moments} exhibits correlations that are substantially larger. This difference stems from systematic differences in the axis ratios resulting from second moments and isophotal contours, which is shown in Fig.~\ref{fig:compare_isophotal2Stokes_compact}. Evidently, axis ratios estimated from second moments are systematically larger than isophotal axis ratios while orientation angles are unbiased. This implies that in Fig.~\ref{fig:LL_correlation_2moments} galaxies are generally considered to be rounder than they actually are, i.e., the inclination angle is misestimated. Given the formalism of \citet{Lee2010}, this bias bents the estimated angular-momentum-orientation vectors into the line-of-sight, thereby feigning these strong correlations. Our scepticism is further raised by the enormous statistical significance of the correlations, which still seems to hold at separations as large as 10Mpc/h. Finally, we note that the background correlation estimated from randomly shuffling the galaxy positions in the sample \citep[cf.][]{Lee2010} is not zero. This suggests the presence of a strong bias, corrupting the correlation estimate of Fig.~\ref{fig:LL_correlation_2moments}.

\subsection{Point-spread function}

\begin{figure}
\includegraphics[width=8cm]{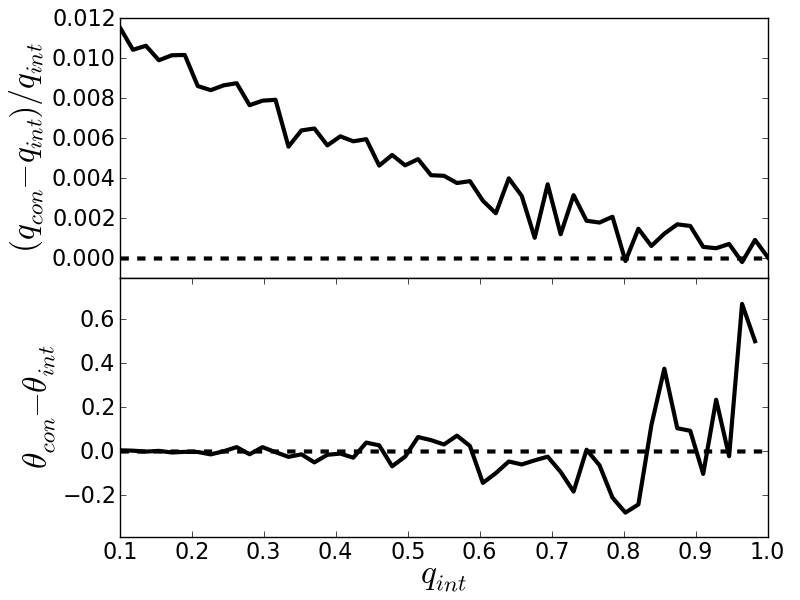}
\caption{Impact of circular Gaussian PSF with Petrosian radius of 1.3 pixel onto convolved axis ratios $q_\textrm{con}$ and orientation angles $\theta_\textrm{con}$ of exponential-disc profiles with Petrosian radii of 15.8 pixels and intrinsic axis ratios $0.1\leq q_\textrm{int}\leq 1$ and orientation angles $\theta_\textrm{int}=30^\circ$. All profiles have been truncated at five scale radii. There was no noise in this simulation. The PSF leads to an overestimation of the axis ratios by at most 1.2\% for highly elongated objects. As the PSF was circular in this test, orientation angles are not affected.}
\label{fig:impact_GaussianPSF_on_discProfile}
\end{figure}

Is this bias an effect of the point-spread function (PSF) which makes galaxies look rounder than they actually are? This is unlikely because all our objects are large compared to the size of the PSF. The median $r$-band Petrosian radius of the 4,211 Scd galaxies with SDSS data is 15.8~pixel, whereas the $r$-band Petrosian radius of the SDSS PSF is approximately 1.3~pixel.\footnote{The $r$-band Petrosian radius of the SDSS PSF has been estimated as the median $r$-band Petrosian radius of 100,000 stars downloaded from the SDSS database.} Consequently, the impact of the PSF should be small. This expectation is supported by Fig.~\ref{fig:impact_GaussianPSF_on_discProfile}, where we simulate the impact of a Gaussian PSF with Petrosian radius 1.3 pixel onto exponential-disc profiles with Petrosian radii of 15.8 pixel and different intrinsic axis ratios. We find a maximum overestimation of axis ratios of only 1.2\%, which is not enough to explain the strong bias in Fig.~\ref{fig:LL_correlation_2moments} or the discrepancy in Fig.~\ref{fig:compare_isophotal2Stokes_compact}.

\subsection{The origin of the bias: Galactic bulges}
\label{sect:galactic_bulges}

\begin{figure}
\includegraphics[width=8cm]{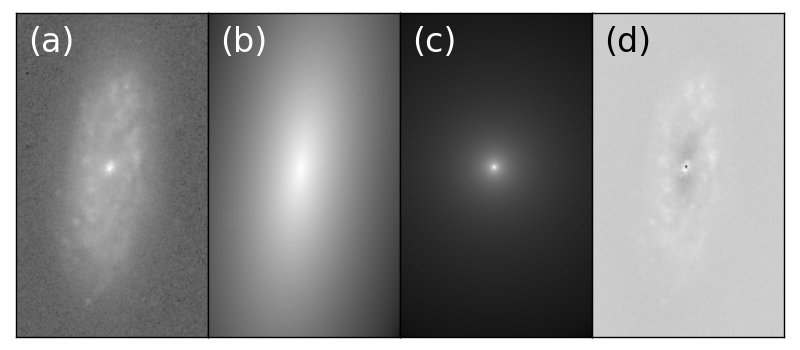}
\caption{Bulge-disc decomposition of an example Scd galaxy ($g$-band). The bulge is a circular deVaucouleur profile, while the disc component is an exponential profile with ellipticity. The bulge is pinned to the pixel of the peak-of-light whereas the centroid of the disc component is free. Panel (a) shows the original galaxy. Panel (b) is the disc component, while panel (c) is the bulge component. Panel (d) displays the fit residuals. The fit was performed by $\chi^2$-minimisation using a Simplex algorithm \citep{Nelder1965} and reached a minimum value of 3.18 per pixel.}
\label{fig:example_bulge_disc_decomposition}
\end{figure}

We are now going to argue that the heavily biased correlation estimate of Fig.~\ref{fig:LL_correlation_2moments} stems from the galactic bulges biasing the second moments and thereby the ellipticity estimates. At first glance, this may seem to be a rather unlikely explanation, since we explicitly selected only Scd galaxies in order to minimise the impact of galactic bulges. However, this hypothesis can explain the substantial discrepancy between isophotal axis ratios and axis ratios based on second moments revealed by Fig.~\ref{fig:compare_isophotal2Stokes_compact}. If bulges were an issue, they would affect the second moments and would lead us to overestimate axis ratios, since bulges are in any case ``roundish''. On the other hand, isophotal ellipticity estimates should be unaffected by the presence of bulges as long as the isophote used is in the disc component. In fact, \citet{Bernstein2010} discuss this issue in the context of shear measurements in weak lensing. We demonstrate that the presence of a bulge can bias the estimate of axis ratio based on second moments. For this purpose, we perform a bulge-disc decomposition of a prototypical Scd galaxy from our data sample, which is shown in Fig.~\ref{fig:example_bulge_disc_decomposition}. Indeed, the axis ratio estimated from the second moments of the complete model (including bulge) is $q_\textrm{b+d}\approx 0.48$, whereas the axis ratio used by the disc model is only $q_\textrm{disc}\approx 0.38$.\footnote{The $g$-band axis ratio noted in the SDSS database for this example galaxy is $q_\textrm{iso}\approx 0.41$ estimated from isophotes and $q_\textrm{mom}\approx 0.63$ estimated from second moments (Stokes parameters). The discrepancy of axis ratios from the SDSS database and the bulge-disc decomposition is the consequence of a non-optimal model.} We conclude that the bulge is well capable of biasing the ellipticity estimate substantially, even in the case of Scd galaxies.

As another test for our hypothesis to pass, we compare the axis ratios based on isophotes and second moments for Sab galaxies from the sample of \citet{HuertasCompany2011}. As Sab galaxies have more prominent bulges than Scd galaxies, we would expect a stronger bias than in Fig.~\ref{fig:compare_isophotal2Stokes_compact}. We select all galaxies with $p_\textrm{Sab}\geq 0.8$ and download the $r$-band Stokes parameters from the SDSS database, if available. For the resulting 8,496 Sab galaxies, Fig.~\ref{fig:compare_isophotal2Stokes_compact} also shows the comparison of ellipticities estimated from isophotes and second moments. Evidently, the second moments are biased, too, and the bias is also more pronounced than for Scd galaxies. This confirms our expectation.

\begin{figure}
\includegraphics[width=8cm]{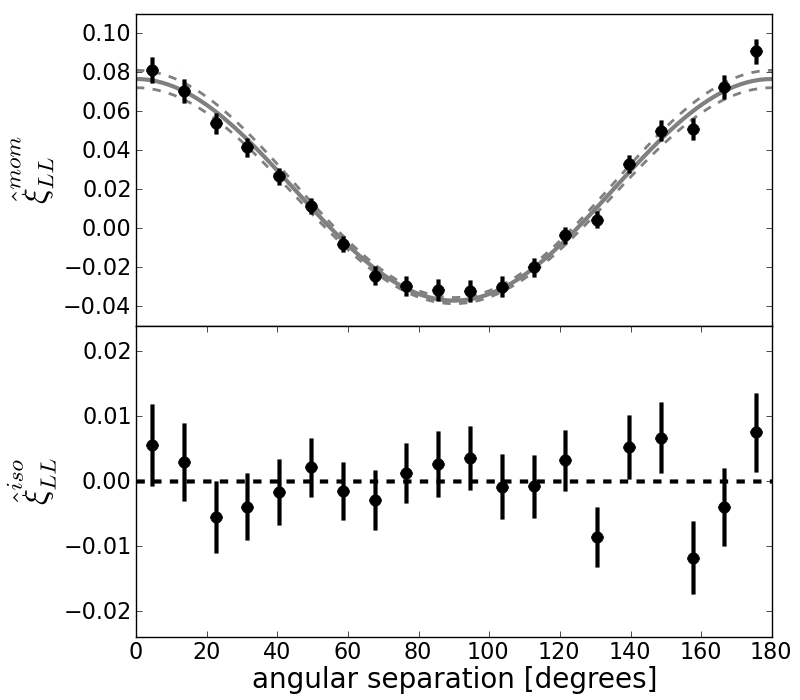}
\caption{Comparing autocorrelations of angular-momentum-orientation vectors in angular separation for ellipticity estimates based on second moments (top) and isophotes (bottom). The bias model of Eq.~(\ref{eq:bias_model_LL}) with $1\sigma$ errors is shown in the top panel.}
\label{fig:comparing_angular_autocorrelations_LL}
\end{figure}

\begin{figure}
\includegraphics[width=8cm]{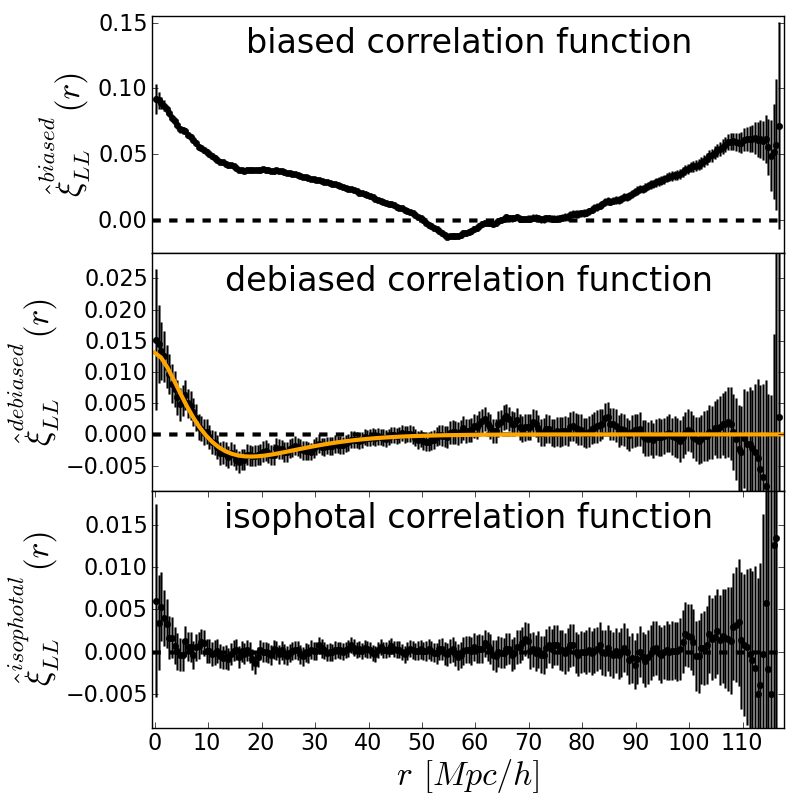}
\caption{Debiasing the autocorrelation function of angular-momentum-orientation vectors. Top panel: The biased autocorrelation function based on ellipticity estimates from second moments. Middle panel: ``Debiased'' correlation function where Eq.~(\ref{eq:bias_model_LL}) has been subtracted from all pairwise projections. The solid orange line is the fit given by Eq.~(\ref{eq:debiased_autocorrelation}). Bottom panel: Autocorrelation function based on isophotal ellipticities.}
\label{fig:debiased_full_correlation_functions_LL}
\end{figure}

From our hypothesis of bulges biasing second moments, we can deduce the following prediction: If galactic bulges indeed bias second moments such that \textit{estimated} angular-momentum-orientation vectors are bent into the line of sight, the \textit{angular} correlation function should exhibit a bias of the form
\begin{equation}\label{eq:bias_model_LL}
b(\theta) = A + B\cos^2\theta \,\textrm{,}
\end{equation}
where $\theta$ now denotes the angular separation of two galaxies.\footnote{The parameter values $A$ and $B$ depend on the details of the bias caused by the galactic bulges and are not generally predictable.} The reason is that due to the bending of orientation vectors, the scalar product $\vec L\cdot\vec L^\prime$ is \textit{on average} equal to the cosine of the two galaxies' separation angle. This prediction is confirmed by Fig.~\ref{fig:comparing_angular_autocorrelations_LL} which strongly suggests that $\hat\xi_\textrm{LL}(\theta)$ is dominated by this bias. This suspect behaviour is also exhibited by the autocorrelation function in real space, as shown in the top panel of Fig.~\ref{fig:debiased_full_correlation_functions_LL}. Figure~\ref{fig:comparing_angular_autocorrelations_LL} also shows that when using isophotal ellipticity estimates, $\hat\xi_\textrm{LL}(\theta)$ does not exhibit such a bias.\footnote{Note that the angular correlation estimate in Fig.~\ref{fig:comparing_angular_autocorrelations_LL} looks worse than the spatial correlation estimate of Fig.~\ref{fig:impact_errors_on_LL_autocorrelation}d. This is due to the fact that the angular correlation function does not use distance information.}

Is it possible to debias the autocorrelation function by subtracting Eq.~(\ref{eq:bias_model_LL}) from all pairwise projections of angular-momentum-orientation vectors? We investigate this question in Fig.~\ref{fig:debiased_full_correlation_functions_LL}, where we show the biased and debiased autocorrelation function. Indeed, the debiased autocorrelation function looks very promising. For later modelling purposes, we parametrise the debiased autocorrelation function by
\begin{equation}\label{eq:debiased_autocorrelation}
\xi_\textrm{LL}(r)\approx (0.013 + 0.002 r - 0.00036 r^2)\exp\left[-\frac{r}{6.1\Mpc/h}\right] \,\textrm{,}
\end{equation}
where no error estimate is required because we only use this fit as input in simulations. Is the debiased autocorrelation function trustworthy? For comparison, Fig.~\ref{fig:debiased_full_correlation_functions_LL} also shows the unbiased autocorrelation function based on isophotal ellipticites. Evidently, the debiased and isophotal autocorrelation functions do not agree. However, this does not necessarily rule out the debiased autocorrelation function because we actually expect that ellipticity estimates based on second moments are less noisy than isophotal ellipticity estimates since they use the whole light distribution instead of a single isophote. Hence, it is not a-priori implausible that the debiased autocorrelation function exhibits more information than the isphotal autocorrelation function.

\begin{figure}
\includegraphics[width=8cm]{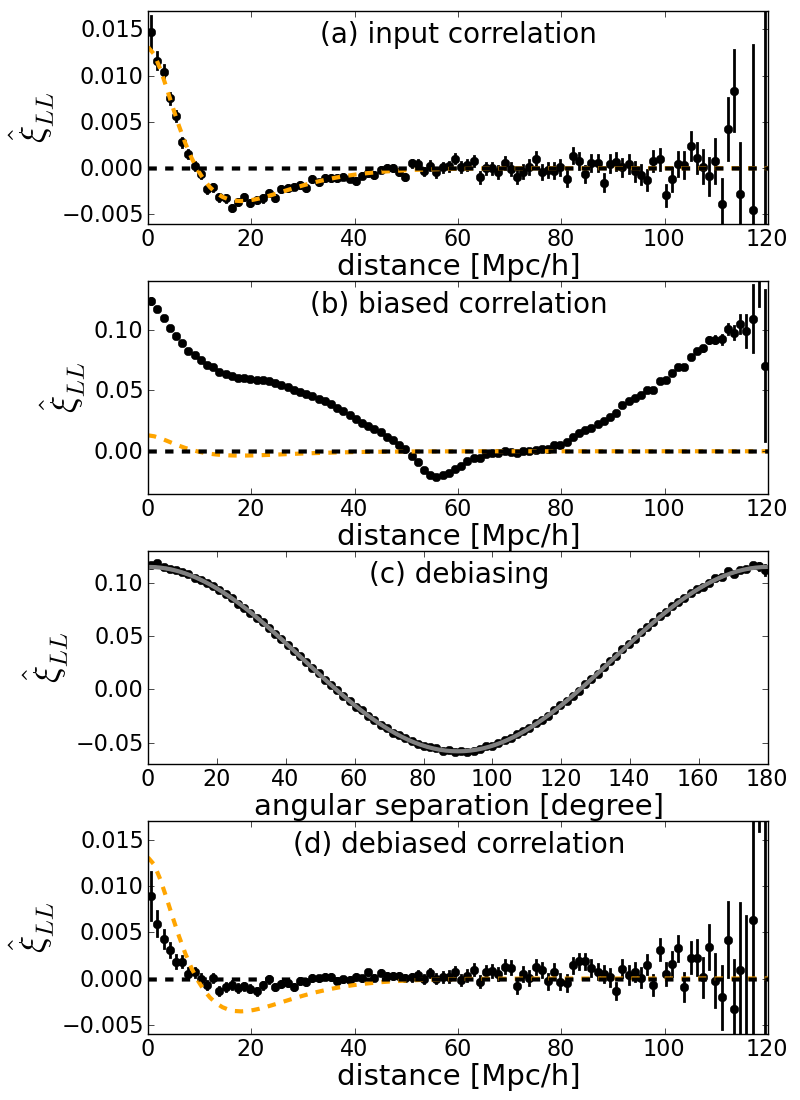}
\caption{Self-consistency test of debiasing the autocorrelation function. Panel (a): The input autocorrelation function as given by Eq.~(\ref{eq:debiased_autocorrelation}), validating our simulation technique. Panel (b): The biased autocorrelation function. Panel (c): The debiasing of the autocorrelation function in angular space. Panel (d): The debiased autocorrelation function, which exhibits significant deviations from the input.}
\label{fig:simulate_debiasing}
\end{figure}

In order to assess the trustworthiness of the debiased autocorrelation estimate, we conduct the following self-consistency test: We take the original galaxies as in Fig.~\ref{fig:debiased_full_correlation_functions_LL}, maintaining their true spatial positions, but when estimating the autocorrelation function, we replace the actual angular-momentum-orientation vectors by simulated vectors which exhibit the correlation function given by Eq.~(\ref{eq:debiased_autocorrelation}). This simulation is described in Appendix~\ref{app:simulating_correlated_LL}. Panel (a) of Fig.~\ref{fig:simulate_debiasing} validates our simulation method. We then simulate the bias of second moments. For every galaxy, we take the simulated angular-momentum-orientation vector and infer the actual axis ratio $q_\textrm{true}$ from it. Motivated by the top left panel of Fig.~\ref{fig:compare_isophotal2Stokes_compact}, we then replace the true axis ratio by an ``overestimate'' drawn from the uniform distribution over the interval $[q_\textrm{true},1]$. Using this biased axis ratio, we recompute the angular-momentum-orientation vector and estimate the correlations. As shown in panel (b) of Fig.~\ref{fig:simulate_debiasing}, the resulting biased autocorrelation function closely resembles the observation from Fig.~\ref{fig:debiased_full_correlation_functions_LL}. For debiasing, we then also estimate the autocorrelation in angular space, as shown in panel (c) of Fig.~\ref{fig:simulate_debiasing}. Indeed, the estimate is dominated by a bias of the form of Eq.~(\ref{eq:debiased_autocorrelation}), i.e., our bias simulation is realistic. We then estimate the debiased autocorrelation function, which is shown in panel (d). Evidently, the debiased result exhibits systematic and significant deviations from the input autocorrelation function. We emphasise that the debiased result is \textit{not} an obscured version of the input correlation function. Neither their difference nor their ratio is a constant, i.e., the debiasing was not successful. Consequently, the debiasing is not self-consistent and the debiased autocorrelation estimate shown in Fig.~\ref{fig:debiased_full_correlation_functions_LL} is \textit{not} trustworthy.

\section{Improvements and potential of future surveys}
\label{sect:future_surveys}

We showed in Figs.~\ref{fig:impact_errors_on_handedness_autocorrelation}d and~\ref{fig:impact_errors_on_LL_autocorrelation}d that with current data there are no statistically significant autocorrelations. What can be done to improve these results? In this section, we briefly elaborate on improvements of ellipticity estimates and the potential of future sky surveys, namely PanSTARRS, LSST and EUCLID, to enhance the estimates of handedness and angular-momentum-orientation autocorrelations. We discuss the impact of number statistics and improvements of redshift estimates. We also discuss morphological classification and estimation of front-edges of disc galaxies.

\subsection{Improving ellipticity estimates}

We demonstrated in Sect.~\ref{sect:eps_biases_2nd_moments} that ellipticity estimates based on second moments are strongly biased by galactic bulges even for Scd galaxies. In fact, Fig.~\ref{fig:comparing_angular_autocorrelations_LL} suggests that correlation estimates based on second moments are completely dominated by this bias which overwrites the desired astrophysical signal. Therefore, we conclude that ellipticity estimates based on second moments overestimate axis ratios and thereby corrupt estimates of angular-momentum-orientation autocorrelation. This bias also corrupts similar correlation estimates, such as ellipticity autocorrelations \citep[e.g.][]{Blazek2011}, leading us to overestimate the impact of disc alignment on weak-lensing studies. What are alternative ellipticity estimators? This same bias also applies to adaptive moments \citep{Bernstein2002,Hirata2003} in this context. Furthermore, model-based ellipticity estimates are problematic, since nearby disc galaxies usually exhibit rich azimuthal structures, which are virtually impossible to model faithfully. The only kind of model designed to describe such rich azimuthal structure are basis-function expansions \citep[e.g.][]{Massey2005,Ngan2009}, which unfortunately suffer from other severe conceptual problems \citep{Melchior2009b,Andrae2011a}. We have to conclude that isophotal ellipticities -- though relying on a somewhat arbitrarily chosen isophote\footnote{The SDSS pipeline uses the 25 magnitudes per square arcsecond isophote.\newline http://www.sdss.org/dr6/algorithms/classify.html\#photo\_stokes} -- are the only useful ellipticity estimates for investigations of angular-momentum-orientation autocorrelation, since they are closest to the disc ellipticity.

There is yet another serious conceptual issue we have to face. In the weak-lensing context galaxies are usually rather small with radii of a few pixels only. In our case, however, we are considering large extended disc galaxies. Disc galaxies usually exhibit substructures such as galactic bars, rings or star-forming regions. In particular, the Scd galaxies considered by \citet{Lee2010} and in this work typically exhibit very open spiral-arm patterns. For such objects there are considerable ellipticity gradients \citep{Bernstein2010} and ``disc ellipticity'' is not a well defined concept anymore. Therefore, it may be helpful to estimate ellipticities in the near infrared regime, where, e.g., star-forming regions are not as prominent as in the optical regime such that disc galaxies look smoother.

\subsection{Improving number statistics}

An obvious strategy to improve estimates of handedness or angular-momentum-orientation autocorrelations is to increase the number of galaxies in the data sample. For instances, SDSS and thereby Galaxy Zoo cover approximately one quarter of the full sky. How would an extension to an (extragalactic) all-sky survey improve the autocorrelation estimates? If we assume identical depth, this areal extension leaves the galaxy density unchanged, it only increases the number of galaxy pairs in all distance bins.

\begin{figure}
\includegraphics[width=8cm]{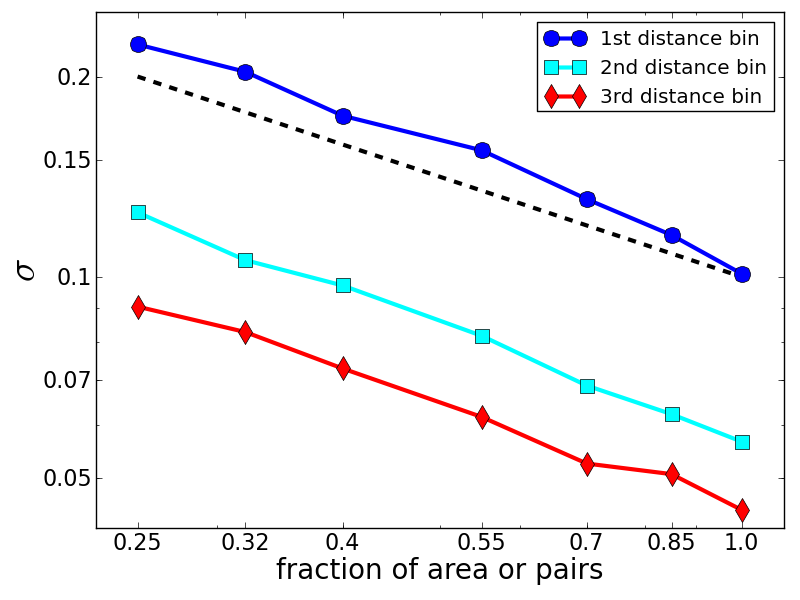}
\caption{Impact of number statistics on the errors of the innermost three distance bins in the marginal handedness autocorrelation function. The $x$-axis shows the fraction of galaxy pairs selected from all pairs, which is equivalent to a survey covering the same fraction of the total survey area. Both axes are in logarithmic scale, i.e., the dependence of the errors is approximately a power law for all three bins. The dashed line indicates a power law of $N^{-1/2}$, where $N$ is the number of pairs in every bin.}
\label{fig:impact_number_stats_on_handedness_errors}
\end{figure}

In order to study the improvement of an enlarged survey area, we randomly draw subsamples from the Galaxy Zoo database (a larger database is not available, so we use smaller databases) and estimate their handedness autocorrelations. In fact, we do not draw the subsamples from the database itself, which would correspond to reducing the galaxy density. Instead, we randomly draw the subsamples from the list of galaxy \textit{pairs}. Figure~\ref{fig:impact_number_stats_on_handedness_errors} clearly shows that the errors in the handedness autocorrelation function are indeed dominated by number statistics, since the errors depend on sample size with a power law of exponent $-\frac{1}{2}$. Consequently, an extension from SDSS to full-sky coverage with SDSS quality would increase the database approximately threefold (the Milky Way obscures roughly one quarter of the sky) and thereby would decrease the errors by a factor of $\sqrt{3}\approx 1.7$. Given the results of Figs.~\ref{fig:impact_errors_on_handedness_autocorrelation}d and~\ref{fig:impact_errors_on_LL_autocorrelation}d, this would clearly be a major break-through in the measurability of potential autocorrelations.

\subsection{Improving redshift estimates}
\label{sect:improving_redshifts}

Reducing the errors in spectroscopic redshift estimates would clearly help in order to reduce the errors in the autocorrelation functions. For instances, the redshift error of $\sigma_z=7.8\cdot 10^{-5}$ at $z = 6.5993\cdot 10^{-2}$ quoted in Fig.~\ref{fig:errors_in_comoving_distance} corresponds to an error in the radial-velocity estimate of $\sigma_v=\frac{c\,\sigma_z}{(1+z)^2}\approx 20.6\kms$. However, given the typical velocity dispersion of galaxies in small groups of $(202\pm 10)\kms$ and in large clusters of $(854\pm 102)\kms$ \citep{Becker2007}, the spectroscopic redshift estimates of SDSS are already picking up peculiar motions of individual galaxies instead of cosmological expansion. Consequently, further improving the accuracy of spectroscopic redshifts cannot improve estimates of, e.g., the handedness autocorrelation function.

Given the impact of uncertainties in spectroscopic redshift estimates on, e.g., the handedness autocorrelation function, it is obvious that \textit{photometric} redshift estimates cannot help to improve the situation. Typically, uncertainties in photometric redshift estimates are two orders of magnitudes larger than uncertainties in spectroscopic redshift estimates. Considering Fig.~\ref{fig:errors_in_comoving_distance}, this would lead to an error in the comoving distance of several tens of Mpc/h. Moreover, though there are many more galaxies with photometric redshift estimates than galaxies with spectroscopic redshift estimates (typically at least one order of magnitude), these additional objects are typically also much fainter because selection for spectroscopic observations is usually triggered by the galaxy's brightness. The faintness of these additional objects would therefore also complicate the morphological classification. For a disc galaxy, the fainter the object, the more difficult it is to identify the disc. Consequently, surveys that offer only photometric but no spectroscopic redshift estimates are of no use to estimate these autocorrelation functions. This essentially rules out PanSTARRS and LSST. Conversely, the EUCLID survey will gather of the order of 100 million spectroscopic redshifts of galaxies. Unfortunately, the galaxy sample observed by EUCLID will have redshifts between 0.5 and 2. As was shown by \citet{Crittenden2001}, estimates of handedness and angular-momentum-orientation correlations are compromised by weak-lensing signals for $z > 0.3$.

\subsection{Morphological classification in future surveys}

Evidently, autocorrelation estimates of handedness and angular-momentum orientation require morphological classification in future surveys. As we cannot probe high-redshift galaxies for this purpose, the morphological classes used by Galaxy Zoo or \citet{HuertasCompany2011} are sufficient and no further diversification is necessary. In particular, this implies that we can build on these two morphological catalogues to classify galaxies in future surveys: First, we match for the galaxies of known morphological types in the new survey. Second, we use the new survey's imaging or spectroscopic data to estimate those galaxy's parameters. Finally, using these parameters and the galaxies of known morphological types as a training sample, we can set up a probabilistic classification algorithm to extend this classification scheme to the new survey catalogue. In fact, this is precisely the same exercise as \citet{HuertasCompany2011} did, but on much larger scale. In particular, the Galaxy Zoo sample with approximately 900,000 visually classified galaxies out to redshift $z\approx 0.5$ would provide an extremely valuable training sample. \citet{Gauci2010} demonstrated that modern classification algorithms perform excellently in reproducing the visual classifications of the Galaxy Zoo sample. This strategy has several advantages: It is easily conductable, it does not require much computational time, and it is highly accurate and objective.

\subsection{Front-edge estimation}
\label{sect:front_edge_estimation}

With so little information in the data, using additional information can be very helpful. Such additional information is provided by an estimate of the disc's front edge, i.e., which edge of the semi-minor axis is pointing towards us. If we can estimate the front-edge, we can use the results as weights $p_a$ and $p_b$ in the correlation estimator of Eq.~(\ref{eq:estimator_LL}). Evidently, if we knew the front edge of every galaxy in our data sample, this would break the geometric degeneracy in the angular-momentum-orientation vector and thereby would improve the correlation estimate.

\subsubsection{Visual classification}

We estimate the front-edge by looking for dust extinction, in particular dust lanes. We visually inspect $g$-band images, since of all five SDSS bands this band is most strongly affected by dust extinction while still being of decent depth. The outcome of such a visual inspection is as follows:
\begin{itemize}
\item Equal weights $p_a=p_b=\frac{1}{2}$ if we are uncertain.
\item Weight of 0.6 to indicate a somewhat uncertain trend.
\item Weight of 0.9 if we believe to be certain.
\end{itemize}
We do not assign a weight of 1 in the last case, since there is always some uncertainty. By construction, this method works best for edge-on discs, since face-on discs do not display dust lanes. Unfortunately, knowing the front-edge would have a larger impact for nearly face-on discs than for edge-on discs \citep[see definitions in][]{Lee2010}. We visually inspected $g$-band images of the 500 largest galaxies, sorted by their Petrosian radii. For smaller galaxies, the resolution is not good enough to identify dust lanes. Unfortunately, we find only very few decisive front-edge classifications, namely 40 Scd galaxies with certain front-edge classifications and 39 with somewhat uncertain front-edge. Consequently, we find no substantial improvement of the marginal correlation estimate. Nevertheless, future sky surveys may have an improved imaging quality, such that a visual front-edge classification is possible for more objects.

\subsubsection{Automated classification}

It is definitely beneficial to obtain a front-edge classification for galaxies with intermediate inclinations, since the rounder the object the larger the information gain. Unfortunately, visual classification via dust lanes is restricted to highly inclined discs. Therefore, the front edge needs to be inferred in a different way, which should ideally be fully automated in order to ensure objectiveness. One potential approach is front-edge classification via colour gradients from dust extinction. However, this requires highly accurate photometric positions. In simple tests, we experienced that already coordinate offsets between the different bands of a tenth of a pixel along the semi-minor axis can compromise such estimates. Another approach is front-edge classification via dust extinction in single-band photometry. In the case of SDSS, this would ideally be the $g$-band, where the impact of dust extinction is larger than in $r,i,z$ whereas the $g$-band is not as shallow as the $u$-band. This approach would compare the fluxes above and below the major axis, thereby estimating the front edge. In contrast to colour-based methods, this approach does not rely on accurate photometric positions. However, like any automated method for front-edge classification, it suffers from several other effects such as star-forming regions in the galaxy or foreground stars which compromise colour gradients and flux differences. These effects are the major obstacles which have to be overcome in order to set up a reliable front-edge classification algorithm.

\section{Discussion and conclusions}
\label{sect:conclusions}

We have shown that when all relevant error sources are taken into account, there are no statistically significant autocorrelations, neither of spiral-arm handedness nor of angular-momentum-orientation vectors of Scd galaxies. Previous estimates \citep{Slosar2009,Lee2010} did not account for these error sources and therefore are \textit{conditional} estimates that underestimated the errors and overestimated statistical significance. Nevertheless, this does not yet falsify the tidal-torque theory for two reasons: First, we indeed see indications for potential autocorrelations, though they are not statistically significant. These indications are consistent with the theoretically predicted correlation length of $1\Mpc/h$. Improving the data might help to test these indications. Second, using a KS-test to analyse the angular-momentum-orientation vectors in the Local Group, the null hypothesis of random orientation yields a $p$-value of 64.8\%, i.e., it cannot be rejected. Therefore, there is no evidence that disc alignment is at work in the Local Group. Third, the tidal-torque theory predicts the alignment for angular momenta of dark-matter haloes and \textit{not} for the disc galaxies residing inside these haloes. For instances, \citet{Bosch2002} find a median misalignment of angular momenta of disc galaxies and their host haloes of $\approx 30^\circ$. Furthermore, even minor mergers can significantly disturb the angular momenta of disc galaxies by transferring orbital angular momentum \citep[e.g.][]{Moster2010}. Conversely, we could speculate whether there is some relaxation process compensating, e.g., for perturbations by mergers. However, we do not want to push this discussion too far because we are wary of turning the tidal-torque theory from an empirical into a ``vampirical'' hypothesis where virtually any observational result can be explained such that an empirical falsification becomes impossible \citep{Gelman2009}.

We must conclude that with currently available SDSS data it is not possible to place decisive constraints on the free parameters of theoretical models. We discussed that already a full-sky survey of SDSS quality might improve the situation such that these autocorrelations could become statistically significant. Furthermore, we argued that photometric redshift estimates of SDSS quality have too large errors to be useful for this task, instead spectroscopic redshift estimates are necessary. Finally, we discussed that a front-edge classification of disc galaxies might improve the autocorrelation estimate of angular-momentum orientation, since it breaks the geometric degeneracy of the galaxy's disc inclination. However, we find that imaging data allows visual front-edge classification only for a minute fraction of objects in the catalogue, whereas automated front-edge classification is severly hampered by foreground stars and star-forming regions. Unfortunately, there are no upcoming surveys that fulfill all these requirements. Consequently, the search for autocorrelations of angular momenta of disc galaxies may remain an open issue for the unforeseeable future.

We demonstrated that ellipticity estimates based on second moments of the galaxies' light distributions are strongly biased by the presence of galactic bulges even for Scd galaxies. This bias corrupts autocorrelation estimates of angular-momentum orientation because it dominates over the expected astrophysical signal. For instances, this leads to an overestimation of the impact of disc alignment in weak-lensing studies \citep{Blazek2011}.

\paragraph*{Acknowledgements}

First and foremost, RA thanks Ellen Andrae for extensive discussions on the contents of this article. RA furthermore thanks Andrea Maccio for discussing the interpretation of the results and Nicolas Martin for valuable information about the Local Group. RA also likes to thank An\v{z}e Slosar who kindly provided his list of Galaxy Zoo object pairs that were visually classified as rogue pairs. Furthermore, Jounghun Lee was very kind in helping us to reproduce her results. Both, An\v{z}e Slosar and Jounghun Lee, also provided detailed comments that were of great help to improve this manuscript. RA also thanks Marc Huertas-Company, who kindly provided additional morphological information on his catalogue of classifications. The authors also want to thank Matthias Bartelmann and Coryn Bailer-Jones, who helped to improve this article. RA is funded by the Klaus Tschira Foundation via the Heidelberg Graduate School of Fundamental Physics (HGSFP). KJ is supported by the Emmy-Noether-programme of the DFG.

\bibliographystyle{aa}

\def\physrep{Phys. Rep.}%
\def\apjs{ApJS}%
\def\apjl{ApJL}%
\def\apj{ApJ}%
\def\aj{AJ}%
\def\aap{A\&A}%
\def\aaps{A\&AS}%
\def\mnras{MNRAS}%
\def\memras{MmRAS}%
\def\araa{ARA\&A}%
\def\jcap{JCAP}
\def\pasa{PASA}
\bibliography{bibliography}

\appendix

\section{Simulating pairs of angular-momentum-orientation vectors}
\label{app:simulating_LL}

In this appendix, we explain how to simulate pairs of angular-momentum-orientation vectors which should exhibit a given correlation.

\subsection{Uncorrelated, orthonormal orientation vectors}
\label{app:simulating_uncorrelated_LL}

As the orientation vectors indicate directions, the samples are drawn from the uniform distributions $\varphi\in[0,2\pi)$ and $\cos\vartheta\in[-1,1]$ of the two polar angles $\varphi$ and $\vartheta$. A random orientation vector is then given by
\begin{equation}
\vec\ell_1 = \left(\begin{array}{c}
\cos\varphi\sin\vartheta \\
\sin\varphi\sin\vartheta \\
\cos\vartheta
\end{array}\right) \;\textrm{.}
\end{equation}
This vector is normalised, i.e., $\vec\ell_1\cdot\vec\ell_1=1$. Sampling a uniform angle $\phi\in[0,2\pi)$, a second random orientation vector is
\begin{equation}
\vec\ell_2 = \sin\phi\left(\begin{array}{c}
-\sin\varphi \\
\cos\varphi \\
0
\end{array}\right) + \cos\phi\left(\begin{array}{c}
\cos\varphi\cos\vartheta \\
\sin\varphi\cos\vartheta \\
-\sin\vartheta
\end{array}\right) \;\textrm{.}
\end{equation}
This vector is again normalised, i.e., $\vec\ell_2\cdot\vec\ell_2=1$, and also orthogonal to the first, i.e., $\vec\ell_1\cdot\vec\ell_2=0$.

\subsection{Pairs of correlated orientation vectors}
\label{app:simulating_correlated_LL}

In the first step, we sample a pair of uncorrelated angular-momentum-orientation vectors $\vec\ell_1$ and $\vec\ell_2$ as described in the previous section. In the second step, we mix these two uncorrelated vectors such that we obtain two correlated vectors,
\begin{equation}
\vec L_a = \cos\alpha\,\vec\ell_1 + \sin\alpha\,\vec\ell_2 \;\textrm{,}
\end{equation}
\begin{equation}
\vec L_a^\prime = \cos\beta\,\vec\ell_1 + \sin\beta\,\vec\ell_2 \;\textrm{,}
\end{equation}
and their counter-parts due to the front-edge degeneracy,
\begin{equation}
\vec L_b = \cos\alpha\left[\vec\ell_1 - 2(\vec e_r\cdot\vec\ell_1)\vec e_r\right] + \sin\alpha\left[\vec\ell_2 - 2(\vec e_r\cdot\vec\ell_2)\vec e_r\right] \;\textrm{,}
\end{equation}
\begin{equation}
\vec L_b^\prime = \cos\beta\left[\vec\ell_1 - 2(\vec e_r^{\,\prime}\cdot\vec\ell_1)\vec e_r^{\,\prime}\right] + \sin\beta\left[\vec\ell_2 - 2(\vec e_r^{\,\prime}\cdot\vec\ell_2)\vec e_r^{\,\prime}\right] \;\textrm{,}
\end{equation}
where $\vec e_r$ and $\vec e_r^{\,\prime}$ are unit vectors pointing from the coordinate origin towards the positions of both galaxies. Due to the orthonormality of $\vec\ell_1$ and $\vec\ell_2$, all these vectors are unit vectors. The two mixing angles $\alpha$ and $\beta$ have to be chosen such that the desired input correlation
\begin{displaymath}
\xi_\textrm{input} = \frac{1}{4}\left( \langle(\vec L_a\cdot\vec L_a^\prime)^2\rangle + \langle(\vec L_a\cdot\vec L_b^\prime)^2\rangle + \langle(\vec L_b\cdot\vec L_a^\prime)^2\rangle\right.
\end{displaymath}
\begin{equation}\label{eq:app:imposed_correlation}
\left. + \langle(\vec L_b\cdot\vec L_b^\prime)^2\rangle \right) - \frac{1}{3}
\end{equation}
is exhibited by the sampled pairs of orientation vectors. This provides only a single constraint, i.e., we are allowed to freely choose one mixing angle. For convenience, we choose $\alpha=0$ such that $\vec L_a = \vec\ell_1$, which simplifies the calculations. We now need to compute the four expectation values.

\subsubsection{Computing the first term}
\label{sect:correlation_LaLa_theo}

We start by computing $\langle(\vec L_a\cdot\vec L_a^\prime)^2\rangle$, which is the simplest term and also presents the basic arithmetic steps. Evidently,
\begin{equation}
\vec L_a\cdot\vec L_a^\prime = \cos\beta\,\vec\ell_1\cdot\vec\ell_1 + \sin\beta\,\vec\ell_1\cdot\vec\ell_2 \;\textrm{.}
\end{equation}
Using $\vec\ell_1\cdot\vec\ell_1=1$ and $\vec\ell_1\cdot\vec\ell_2=0$, this expression simplifies to
\begin{equation}
\vec L_a\cdot\vec L_a^\prime = \cos\beta \;\textrm{.}
\end{equation}
The autocorrelation is then given by
\begin{equation}\label{eq:correlation_LaLa_theo}
\langle(\vec L_a\cdot\vec L_a^\prime)^2\rangle = \cos^2\beta \;\textrm{.}
\end{equation}

\subsubsection{Computing the other terms}

The other three terms in Eq.~(\ref{eq:app:imposed_correlation}) are computed in precisely the same way. We obtain
\begin{equation}\label{eq:correlation_LaLb_theo}
\langle(\vec L_a\cdot\vec L_b^\prime)^2\rangle = \frac{7}{15}\cos^2\beta + \frac{4}{15}\sin^2\beta \;\textrm{.}
\end{equation}
As the correlation estimate is invariant under exchanging the pair, we can directly conclude that
\begin{equation}\label{eq:correlation_LbLa_theo}
\langle(\vec L_b\cdot\vec L_a^\prime)^2\rangle = \frac{7}{15}\cos^2\beta + \frac{4}{15}\sin^2\beta \;\textrm{,}
\end{equation}
as well. The final term is given by
\begin{displaymath}
\langle(\vec L_b\cdot\vec L_b^\prime)^2\rangle = \left(\frac{7}{15} - \frac{8}{5}(\vec e_r\cdot\vec e_r^{\,\prime})^2 + \frac{32}{15}(\vec e_r\cdot\vec e_r^{\,\prime})^4\right)\cos^2\beta
\end{displaymath}
\begin{equation}\label{eq:correlation_LbLb_theo}
 + \left(\frac{4}{15} + \frac{4}{5}(\vec e_r\cdot\vec e_r^{\,\prime})^2 - \frac{16}{15}(\vec e_r\cdot\vec e_r^{\,\prime})^4\right)\sin^2\beta \;\textrm{,}
\end{equation}
which depends on the angular separation $\vec e_r\cdot\vec e_r^{\,\prime}$ of the galaxy pair that is simulated. This dependence is inherited from flipping the radial component of \textit{both} angular-momentum-orientation vectors due to an unknown front edge.

\subsubsection{Mixing angle}

Inserting all four terms into Eq.~(\ref{eq:app:imposed_correlation}), we can solve for the mixing angle for a given input correlation. The result is
\begin{equation}
\cos\beta = \sqrt{\frac{1}{3} + \frac{20\xi_\textrm{input}}{16(\vec e_r\cdot\vec e_r^{\,\prime})^4-12(\vec e_r\cdot\vec e_r^{\,\prime})^2+8}} \,\textrm{.}
\end{equation}
This mixing angle is used in Sect.~\ref{sect:galactic_bulges}.

\bsp

\label{lastpage}

\end{document}